\documentclass[reprint,superscriptaddress, nofootinbib,longbibliography,twocolumn]{revtex4-2}
\pdfoutput=1
\usepackage{graphicx}
\usepackage{amssymb}
\usepackage{lipsum}
\usepackage[linktoc=none]{hyperref}
\usepackage[dvipsnames]{xcolor}
\usepackage{dcolumn}
\usepackage[english]{babel}
\usepackage[utf8x]{inputenc}
\usepackage[T1]{fontenc}
\usepackage{eqnarray,amsmath,physics,bbold}
\usepackage{ dsfont }
\usepackage[title]{appendix}
\usepackage{xcolor}
\usepackage{breqn}

\usepackage{hyperref}
    \hypersetup{
    colorlinks   = true,    
    urlcolor     = blue,    
    linkcolor    = blue,    
    citecolor    = red      
    }

\DeclareMathAlphabet{\mathpzc}{OT1}{pzc}{m}{it} 

\newcommand{\nmean}[1]{\langle n_{#1} \rangle}

\newcommand{\vek}{\Vec{k}}

\newcommand{\INFN}{INFN - Sezione di Napoli, Complesso Univ. Monte S. Angelo, I-80126 Napoli, Italy}
\newcommand{\UNINA}{Physics Department "Ettore Pancini", Universit\'a degli studi di Napoli "Federico II", Complesso Univ. Monte S. Angelo, Via Cintia, I-80126 Napoli, Italy}
\newcommand{\UNISA}{Physics Department "E.R. Caianiello", Universit\'a degli studi di Salerno, Via Giovanni Paolo II, 132, I-84084 Fisciano (Sa), Italy}
\newcommand{\CNR}{CNR-SPIN Napoli Unit, Complesso Univ. Monte S. Angelo, Via Cintia, I-80126 Napoli, Italy}

\begin{document}
\title{Gate tunable anomalous Hall effect: a Berry curvature probe at oxides interfaces}

\author{M. Trama}
\email{mtrama@unisa.it}
\affiliation{\UNISA}
\affiliation{\INFN}

\author{V. Cataudella}
\affiliation{\UNINA}
\affiliation{\CNR}

\author{C. A. Perroni}
\affiliation{\UNINA}
\affiliation{\CNR}

\author{F. Romeo}
\affiliation{\UNISA}

\author{R. Citro}
\email{rocitro@unisa.it}
\affiliation{\UNISA}
\affiliation{\INFN}

\begin{abstract}
    The characterization and the experimental measurement of the Berry curvature in solids have become an increasingly relevant task in condensed matter physics. We present the theoretical prediction of a gate tunable anomalous Hall effect (AHE) in a non magnetic oxide interface as a hallmark of a non-trivial Berry curvature. The observed AHE at low-temperatures in the presence of a planar magnetic field comes from a multiband low-energy model with a generalized Rashba interaction that supports characteristic out-of-plane spin and orbital textures. We also discuss strategies for reconstructing the Berry curvature from the AHE non-linearities in (111) SrTiO$_3$ heterostructure interfaces.
\end{abstract}

\maketitle

\section{Introduction}\label{sec:intro}
The anomalous Hall effect (AHE), a hallmark of broken time-reversal symmetry and spin-orbit coupling (SOC), has long been an intriguing even though controversial subject. In fact, some of the theories explain the AHE as an effect of the skew scattering~\cite{skewsc} or side jump mechanism~\cite{sidejump} and goes under the name of {\it extrinsic} AHE. Moreover, several studies have pointed out the intrinsic origin of the AHE,
related to the Berry curvature of the quasiparticles on the Fermi surface~\cite{AHE1,AHE2,AHE3,AHE4,AHE5,AHE6}: the mobile charge carriers gain a transverse momentum due to a magnetic polarization coming from the Berry curvature of the occupied Bloch wave functions (cfr. Eq.~(\ref{intrinsic_sigma})). The corresponding Hall voltage, i.e. anomalous Hall component, can be observed as an additional contribution in the Hall
measurements, superposed on the ordinary Hall effect. This Berry scenario of the AHE has recently attracted interest for its dissipationless and topological nature, and specially because it offers a direct measure of the Berry curvature~\cite{PhysRevLett.126.256601,lesne2022designing,PhysRevResearch.3.L012006}. However, up to now, the experimental realization of AHE in non-magnetic systems and in the moderately dirty limit remains elusive. \\
In this work we propose oxide interfaces— artificially created structures involving transition-metal oxide compounds- as a platform for intrinsic AHE. Here the large SOC and the multi-orbital character of the bands enrich the variety of emergent SOC phenomena~\cite{hwang2012emergent}, like the generation and control of spin and orbital textures~\cite{gariglio2018spin} at the origin of the AHE. In particular, we show the latter as a direct probe of the Berry curvature of the system.
Among the oxides interfaces, recent attention has been devoted to the (111) LAO/STO interface,  whose trigonal geometry leads to exotic behaviours, including analogy with graphene in the conducting state~\cite{doennig2013massive,trama2021straininduced} and anisotropic magnetic transport both in normal and superconducting state~\cite{monteiro2017two,davis2017magnetoresistance}. 
\\Here, we predict a gate tunable intrinsic anomalous Hall conductivity (AHC) coming from out-of-plane spin and orbital textures~\cite{PhysRevLett.120.266802} as a direct probe of a non-vanishing Berry curvature in the presence of an in-plane magnetic field. We show that this behavior arises from a generalized Rashba coupling involving the total angular momentum of the bands, and construct an effective Hamiltonian for the low-filling region, underlining the strong orbital character of the Rashba interaction in the (111) direction. This behavior has not been predicted in the more common (001) LAO/STO, establishing the (111) STO interfaces among the reconfigurable platforms for spin-orbitronics~\cite{bibes}.
\\The plan of the manuscript is the following: in Section~\ref{sec:model} we describe the tight-binding (TB) Hamiltonian used for our analysis. We show that this Hamiltonian gives rise to complex spin and orbital angular momentum patterns throughout the Brillouin zone. In Section~\ref{sec:anom_trans} we show that this Hamiltonian, coupled with an in- or an out-of-plane external magnetic field, induces a non-trivial Berry curvature, and that this is responsible for a non-vanishing AHC. We also discuss the temperature dependence of this conductance. Finally, in Section~\ref{sec:concl} we discuss our results and draw our conclusions.
\section{Model}\label{sec:model}
\begin{figure}
    \centering
    \includegraphics[width=0.49\textwidth]{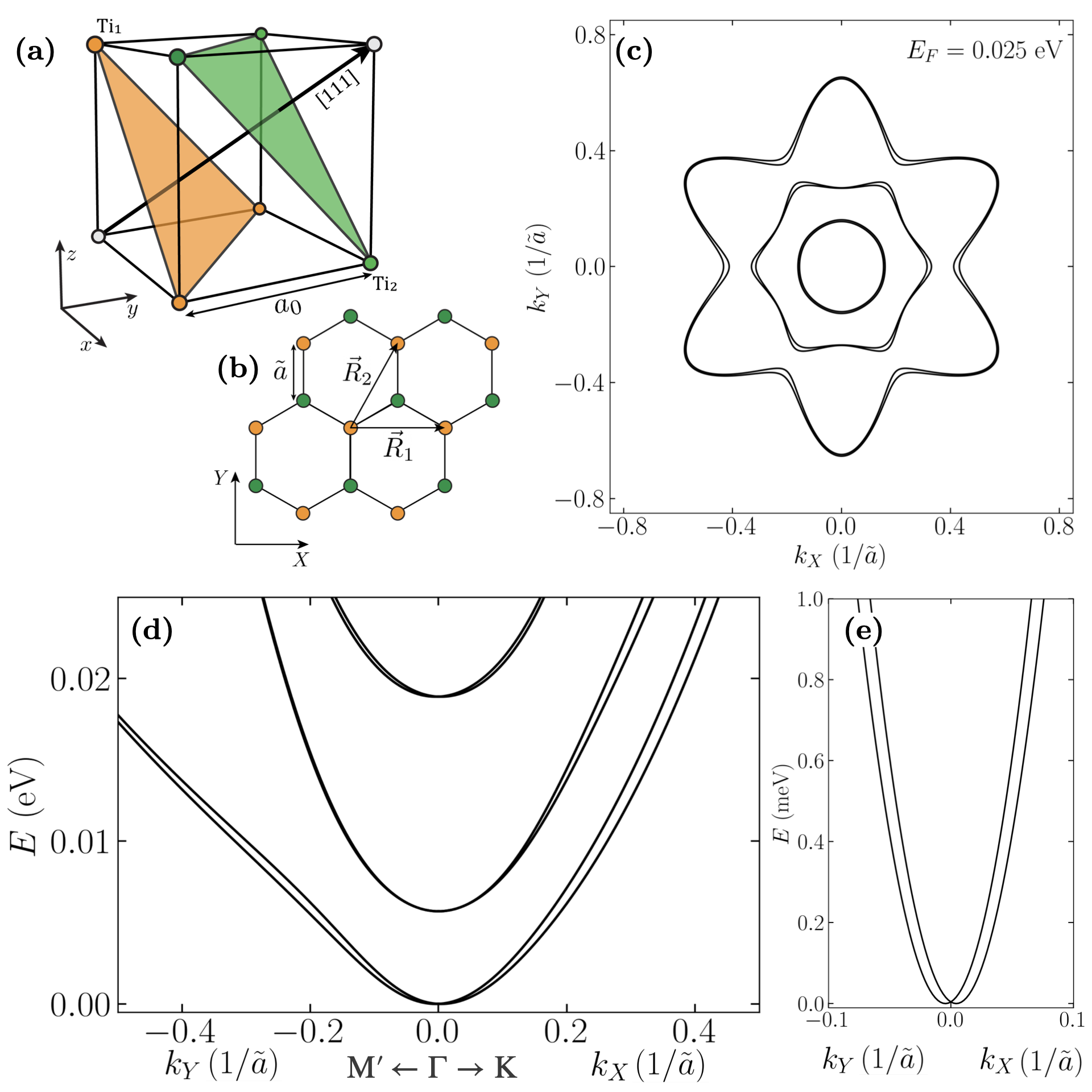}
    \caption{(a) Ti atoms in STO lattice, whose reticular constant is $a_0=0.3905$ nm. The orange and green dots represent atoms belonging to two non-equivalent planes.
    (b) Projection of the two non-equivalent planes of Ti over the (111) plane with our choice of primitive vectors $\Vec{R}_1$ and $\Vec{R}_2$ and $\Tilde{a}=\sqrt{2/3}a_0$. The 3D Brillouin zone and its 2D projection along the (111) direction can be seen in~\cite{PhysRevB.91.075432}. (c) Fermi surface for the Fermi energy $E_F=0.025$ eV.  (d) Band structure along two different directions in the Brillouin zone. (e) Detail of the splitting at low fillings of the first doublet, exhibiting the typical Rashba behaviour. Panels (c) and (d) and (e) are evaluated at $\vec{B}=0$.}
    \label{fig:band_structure}
\end{figure}

We develop a TB model of the electrons at the interface occupying the $t_{2g}$ orbitals of a bilayer of Ti atoms in STO~\cite{Keppler1998}. The Ti lattice projected along the (111) direction is a honeycomb lattice (see Fig.~\ref{fig:band_structure}(a-b)).
The Hamiltonian originating from the three $t_{2g}$ orbitals of the Ti-atoms in the  $\alpha$ and $\beta$ layers reads~\cite{xiao2011interface,trama2021straininduced,trama2022tunable}
\begin{equation}
    H=H_{\text{TB}}+H_{\text{SOC}}+H_{\text{TRI}}+H_{v},
    \label{eq:full_ham}
\end{equation}
where $H_{\text{TB}}$ 
is the hopping Hamiltonian which in $\vek$-space can be written as:
\begin{equation}
    H_{\text{TB}}=\sum_{\vek}\sum_{i,\alpha\beta,\sigma}t_i^{\alpha\beta}(t_D,t_I,\Vec{k}) d_{i\alpha\sigma,\vek}^{\dagger} d_{i\beta\sigma,\vek}
    \label{TBk},
\end{equation}
where $d_{i\alpha \sigma,\vek}$ is
the annihilation operator of
the electron occupying the orbital
$i = {xy, yz, zx}$ belonging to the layer $\alpha,\beta= {\text{Ti}_1, \text{Ti}_2}$ and of spin $\sigma= \pm 1/2$. It depends on the 2D dimensionless quasi-momentum $\vek=\Tilde{a}\Vec{K}$,
where $\Vec{K}$  is the quasi-momentum, $\tilde{a}=\sqrt{\frac{2}{3}}a_0$ is the in-plane lattice constant with $a_0=0.3905$~nm. The matrix $t_i^{\alpha\beta}(t_D,t_I,\Vec{k})$ includes the nearest-neighbour hopping parameters which have been separated in direct $t_D$ and indirect $t_I$ contribution, fixed to the values $t_D=0.5$~eV and $t_I=0.04$~eV~\cite{trama2021straininduced}.
The matrix $t_i^{\alpha\beta}$ has the following form in the basis $\{d_{yz},d_{zx},d_{xy}\}\otimes\{\text{Ti}_1,\text{Ti}_2\}$
\begin{equation}
    t_i^{\alpha\beta}=\begin{pmatrix}
        0 & 0 & 0 & \epsilon_{yz} & 0 & 0 \\
        0 & 0 & 0 & 0 & \epsilon_{zx} & 0 \\
        0 & 0 &0 & 0 & 0 & \epsilon_{xy} \\
        \epsilon_{yz}^* & 0 & 0 & 0 & 0 & 0 \\
        0 & \epsilon_{zx}^* & 0 & 0 & 0 & 0 \\
        0 & 0 & \epsilon_{xy}^* & 0 & 0 & 0 \\
    \end{pmatrix},
\end{equation}
where the interlayer contributions are
\begin{eqnarray}
     \nonumber &\epsilon_{yz}&=-t_D\left(1+e^{i(\frac{\sqrt{3}}{2}k_X-\frac{3}{2}k_Y)}\right)-t_Ie^{-i(\frac{\sqrt{3}}{2}k_X+\frac{3}{2}k_Y)},\\
     &\epsilon_{zx}&=-t_D\left(1+e^{-i(\frac{\sqrt{3}}{2}k_X+\frac{3}{2}k_Y)}\right)-t_Ie^{i(\frac{\sqrt{3}}{2}k_X-\frac{3}{2}k_Y)},\\
     \nonumber &\epsilon_{xy}&=-2t_D\cos(\frac{\sqrt{3}}{2}k_X)e^{-i\frac{3}{2}k_Y}-t_I,
     \label{interlayer}
\end{eqnarray}
where $\vek=k_X \hat{u}_{\overline{1}10}+k_Y\hat{u}_{\overline{1}\overline{1}2}$.
$H_{\rm{SOC}}$ is the atomic SOC coupling, which has the following expression
\begin{equation}
    H_{\text{SOC}}=\frac{\lambda}{2}\sum_{\vek}\sum_{ijk,\alpha,\sigma\sigma'}i\varepsilon_{ijk}
    d_{i\alpha\sigma,\vek}^{\dagger} \sigma^{k}_{\sigma\sigma'}d_{j\alpha\sigma',\vek}
    \label{eq:spinorbit}
\end{equation}
where $\varepsilon_{ijk}$ is the Levi-Civita tensor, and $\sigma^k$ are the Pauli matrices. We fix the SOC coupling $\lambda=0.01$~eV, as a typical order of magnitude~\cite{monteiro2019band}.
$H_{\rm{TRI}}$ is the trigonal crystal field and it takes into account the strain at the interface along the (111) direction. The physical origin of this strain is the possible contraction or dilatation of the crystalline planes along the (111) direction. This coupling has the form~\cite{khomskii2014transition}
\begin{equation}
    H_{\text{TRI}}=\frac{\Delta}{2}\sum_{\vek}\sum_{i\neq j,\alpha,\sigma} d_{i\alpha\sigma,\vek}^{\dagger} d_{j\alpha\sigma,\vek}.
    \label{eq:trigonal}
\end{equation}
We fix $\Delta=-0.005$~eV as reported in~\cite{de2018symmetry}. Finally the last term $H_v$ describes an electric field in the (111) direction, orthogonal to the interface,
which breaks the reflection symmetry.
Differently from previous studies~\cite{xiao2011interface,monteiro2019band}, we treat the effect of the electric field $\vec{E}$ on the orbitals perturbatively. In particular, following Ref.~\cite{shanavas2014theoretical}, we consider that $\vec{E}$ induces an hybridization of the atomic $d$ orbitals of the STO from $|d\rangle$ to $|d^\prime\rangle=c_1|p\rangle+c_2|d\rangle+c_3|f\rangle$, with $|x\rangle$ the atomic orbital $x$, leading to renormalized hopping terms odd in lattice momentum $\vek$.
It can thus be written as
\begin{equation}
    H_v=\frac{v}{2}\sum_{i,\alpha,\sigma,\vek}\xi_{\alpha} d_{i\alpha\sigma,\vek}^{\dagger} d_{i\alpha\sigma,\vek}+
   \sum_{\vek}\sum_{ij,\alpha\beta,\sigma}h_{ij,\vek}^{\alpha\beta}(v) d_{i\alpha\sigma,\vek}^{\dagger} d_{j\beta\sigma,\vek}
    \label{electric}
\end{equation}
where $\xi_{\text{Ti}_1/\text{Ti}_2}=\pm 1$ and the expression of matrix elements $h_{ij,\vek}^{\alpha\beta}(v)$ are obtained in appendix~\ref{Appendice-Electric}. The electric field has been fixed at the value $v=0.2$~eV by comparison with the Rashba splitting evaluated in Ref.~\cite{monteiro2019band}.

In addition to the previous terms in Hamiltonian~(\ref{eq:full_ham}), one can consider the Hamiltonian $H_{\rm{H}}$,
describing the intra-orbital and inter-orbital Hubbard and Hund's interaction~\cite{perroni2007exact}, which has in real space the form
\begin{equation}
    H_{\rm{H}}=U\sum_{r,\alpha}n_{r\alpha\uparrow}n_{r\alpha\downarrow}+\frac{1}{2}\sum_{r,\alpha\neq\beta,\sigma\sigma^{\prime}}(U^{\prime}-J\delta_{\sigma\sigma^{\prime}})n_{r\alpha\sigma}n_{r\beta\sigma^{\prime}},
    \label{hund-text}
\end{equation}
where, $n_{r\alpha\sigma}$ is the number operator of the state located in the site $r$.
For the regime of small interaction parameters $(U,U^\prime,J)$,
we have verified that the effect of Hamiltonian~(\ref{hund-text}) is simply a renormalization of the chemical potential. The analysis is given in appendix~\ref{Appendice-Correlazioni}. Therefore in the following we choose $U=0$, $J=0$ and $U^\prime=0$. Moreover, one can include a tetragonal distortion at low temperature \cite{pai2018physics} through a contribution $H_{\zeta}$. This is further analyzed in the appendix~\ref{Appendice-tetra}.\\
The Hamiltonian~(\ref{eq:full_ham}) leads to the band structure in Fig.~\ref{fig:band_structure}(d), where a Rashba-like splitting appears in the lowest band due to the interplay between SOC and reflection symmetry breaking term. The splitting between the Kramers doublets, visible in the Fermi surface in Fig.~\ref{fig:band_structure}(c), is consistent with previous results~\cite{he2018observation}.
The behaviour for low filling can be recovered by the effective Hamiltonian derived from a perturbative approach truncated at the order $\sim O(\vek^2)$:
\begin{equation}
     H_{\rm{eff}}(\vek)=\sum_{i}\mathcal{E}_{i \vek} (\mathbb{1}-L_i^2)-\frac{\lambda}{2} \hat{L}\cdot\hat{S}-\frac{3\Delta}{2} L_{111}^2+\mathcal{F}(\vek\times\hat{L})+E_0,
     \label{lsmodel}
\end{equation}
where  $\mathcal{E}_i(\vek)$ 
is the renormalized band dispersion coming from the second order expansion, $L_i$ and $S_i$ are the $i$-th component of the orbital and spin angular momentum operator for $L=1$ and $S=1/2$, respectively; the second term is a spin-orbit interaction; $L_{111}$ is the projection of $\vec{L}$ along the $(111)$ direction and the second to last term is an orbital-Rashba interaction~\cite{revieworbitronic2021}, whose coefficient $\mathcal{F}=0.0035$~eV and it is derived in appendix~\ref{Appendice-Electric}. We have verified that the effective Hamiltonian gives the same spectrum of Hamiltonian~(\ref{eq:full_ham}) up to $|\vek|\sim 0.5$ (in unit of $\Tilde{a}$) while all the results are obtained using the full Hamiltonian. In Eq.~(\ref{lsmodel}) a first signature of a generalized Rashba coupling comes from the term $(\vek\times\hat{L})$.
Near $\vek=0$, the Hamiltonian is dominated by atomic SOC, so we can use as a basis the eigenstates of $\Vec{J}=\Vec{S}+\Vec{L}$. Evaluating the Hamiltonian over a multiplet of definite total angular momentum $J$, the effective Hamiltonian can be written as
\begin{equation}
     H_{\rm{eff}}(|\vek|\sim0)=-\tilde{\Delta} J_{111}^2+\tilde{\mathcal{F}}(\vek\times\hat{J})+\sum_{i}\mathcal{E}_i(\vek) (\nu\mathbb{1}-\tau J_i^2),
    \label{jmodel}
\end{equation}
where $J_{111}$ is the projection of the total angular momentum in the (111) direction, $\nu$, $\tau$ are the Landé factors, and $\tilde{\Delta}$ and $\tilde{\mathcal{F}}$ are rescaled as well by the Landé fatcors. The SOC splits the bands in two groups: four states of lower energy with $J=3/2$ and a doublet of $J=1/2$ at higher energy. When we evaluate this operators on the lower quadruplet states we obtain $\nu=3/4$, $\tau=1/3$, $\tilde{\Delta}=\Delta/2$ and $\tilde{\mathcal{F}}=2\mathcal{F}/3$.
Since the trigonal crystal field splits the states according to their value of $\Vec{J}$ along the $(111)$ direction, the first doublet in Fig.~\ref{fig:band_structure} has $J_{111}=\pm 1/2$, while the second doublet has $J_{111}=3/2$. 
For this reason the splitting of the second doublet depicted in Fig~\ref{fig:band_structure}(d) is not linear in the wavevector $\vec{k}$, as for the canonical Rashba interaction, but cubic~\cite{notaJ}. This cubic splitting is naturally encoded in our multiband model giving rise to the so called warping term, relevant for the AHC discussed below~\cite{lesne2022designing}.
We also note that Eq.~(\ref{jmodel}) is not valid for a trigonal crystal field comparable or even larger than the SOC, since the eigenstates of total angular momentum are not anymore a reasonable approximation to the real eigenstates of the problem. 
\\The interplay between spin and orbital angular momentum causes peculiar patterns shown in Fig.~\ref{patterns} (see appendix~\ref{Appendice-totAng} for the total angular momentum patterns).
\begin{figure}
    \centering
    \includegraphics[width=0.48\textwidth]{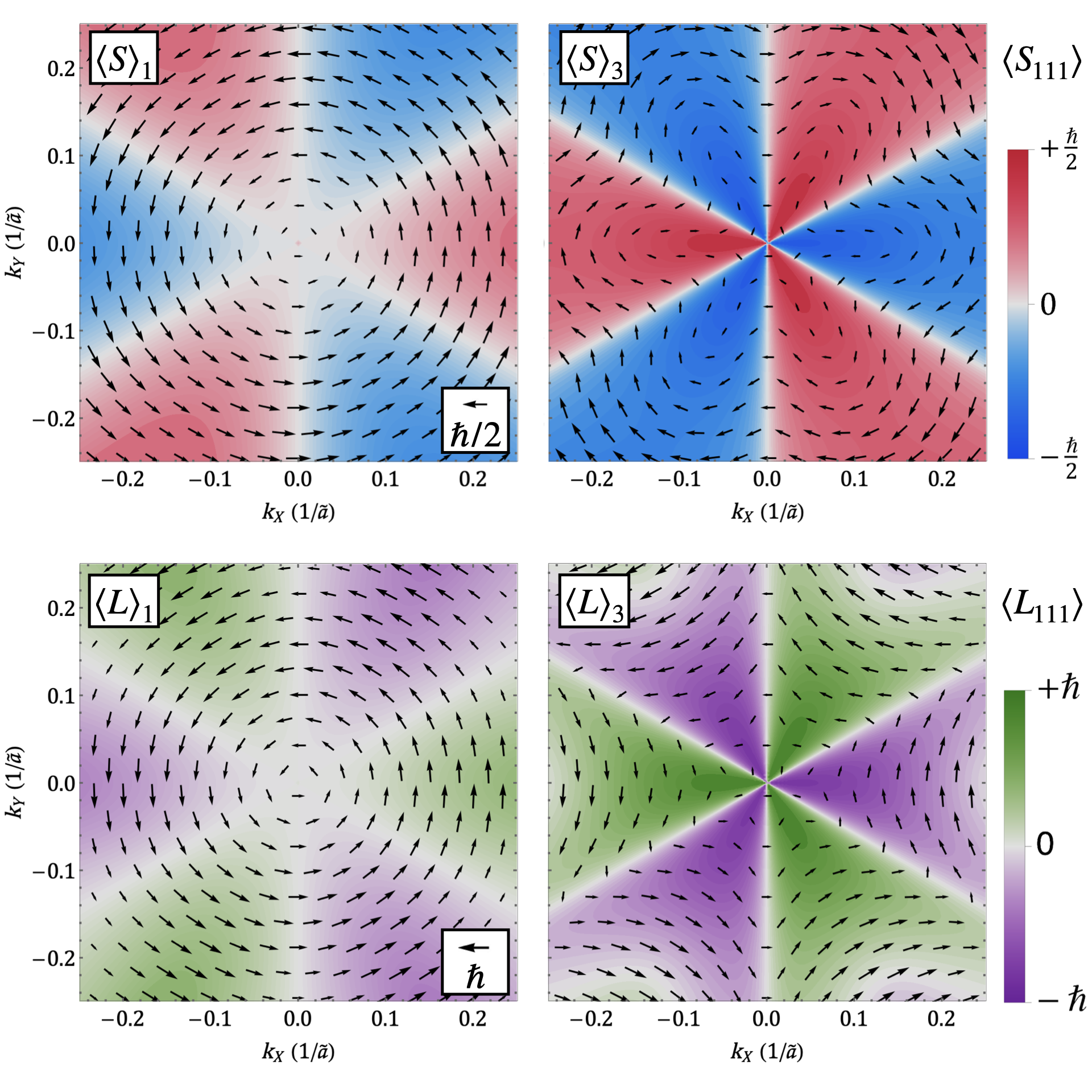}
    \caption{In- and out-of-plane spin and angular momentum modulation for the first and the third band. The second and the fourth are specular to the ones shown. The in-plane patterns are represented by the arrows and are obtained by computing the mean value of the spin (orbital) components $S_{[\overline{1}10]}$ ($L_{[\overline{1}10]}$) and the $S_{[\overline{11}2]}$ ($L_{[\overline{1}10]}$) over the eigenstates of the chosen band. The colours indicate the out-of-plane modulation.}
    \label{patterns}
\end{figure}
For the first two bands the pattern is Rashba-like with a spin amplitude changing with $\vek$. The third and the fourth bands have a Rashba-like vortex near the $\Gamma$ point, replaced by six secondary vortices at larger $\vek$. In both cases an out-of-plane component is present because of the quasi-2D nature of the system.

\section{Anomalous transport property}\label{sec:anom_trans}
\begin{figure}
    \centering
    \includegraphics[width=0.48\textwidth]{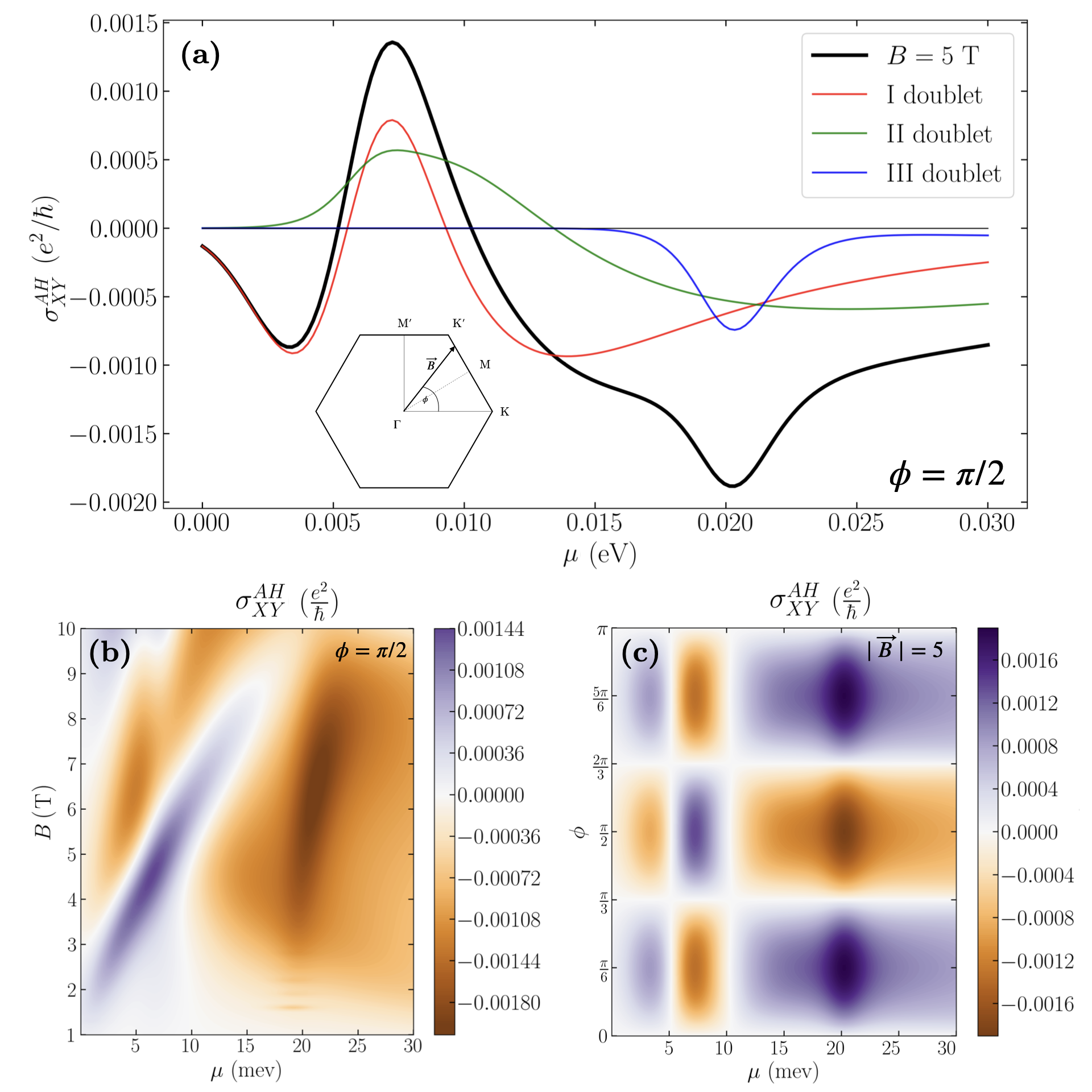}
    \caption{
    (a) AHC as a function of the chemical potential $\mu$ for an in-plane magnetic field $|B|=5$ T, with an in-plane angle $\phi=\pi/2$ at a temperature $T=10$ K.  The colors indicate the sum of the contributes over two specific bands.
    (b) AHC as a function of $\mu$ and for different values of $|\Vec{B}|$ with $\phi=\pi/2$ and $T=10$ K. 
    (c) AHC as a function of $\mu$ for different values of $\phi$ at $|\Vec{B}|=5$ T and $T=10$ K.}
    \label{contour_inplane}
\end{figure}
The Hamiltonian described in Section~\ref{sec:model} possesses time-reversal invariance, under which the AHC vanishes identically.
If we add to the Hamiltonian~(\ref{eq:full_ham}) a Zeeman term $H_B$ of the form
\begin{equation}
    H_B=-\mu_{B}\Vec{B}\cdot(\hat{L}+2\hat{S})
\end{equation}
the time-reversal symmetry of the model is broken. This leads to a non-vanishing out-of-plane Berry curvature.  To evaluate its effect on the intrinsic AHE, we employ a semi-classical approach based on the Boltzmann equations within the time-relaxation approximation.
In the presence of a small in-plane electric field $\Vec{E}_{\rm{ext}}$, the electron group velocity is given by:
\begin{equation}
    \Vec{v}_g=\frac{1}{\hbar}\frac{\partial\varepsilon}{\partial\vek}-\frac{e}{\hbar}\Vec{E}_{\rm{ext}}\times\Vec{\Omega},
\end{equation}
where the former term is the standard dynamical contribution, while the latter is the geometric contribution connected to the Berry curvature $\Vec{\Omega}$. This term gives origin to the intrinsic AHE, mainly discussed in ferromagnetic systems~\cite{RevModPhys.82.1539}:
\begin{equation}
    \sigma_{XY}^{\text{AH}}=-
    \frac{e^2}{\hbar}\sum_{n\hspace{0.05cm}occ}\int_{BZ}\Omega_{n\vek} f_{th,\vek}(T)\hspace{0.1cm}\frac{d^2\vek}{(2\pi)^2},
    \label{intrinsic_sigma}
\end{equation}
where $f_{th}$ is the Fermi distribution at temperature $T$ (where $X$ and $Y$ are the ($\overline{1}10)$ and the ($\overline{11}2)$ directions) and $\Omega_{n\vek}$ refers to the $(111)$ component of the Berry curvature of the n-th band, the only non-vanishing component for a 2D system.
\begin{figure}
    \centering
    \includegraphics[width=0.45\textwidth]{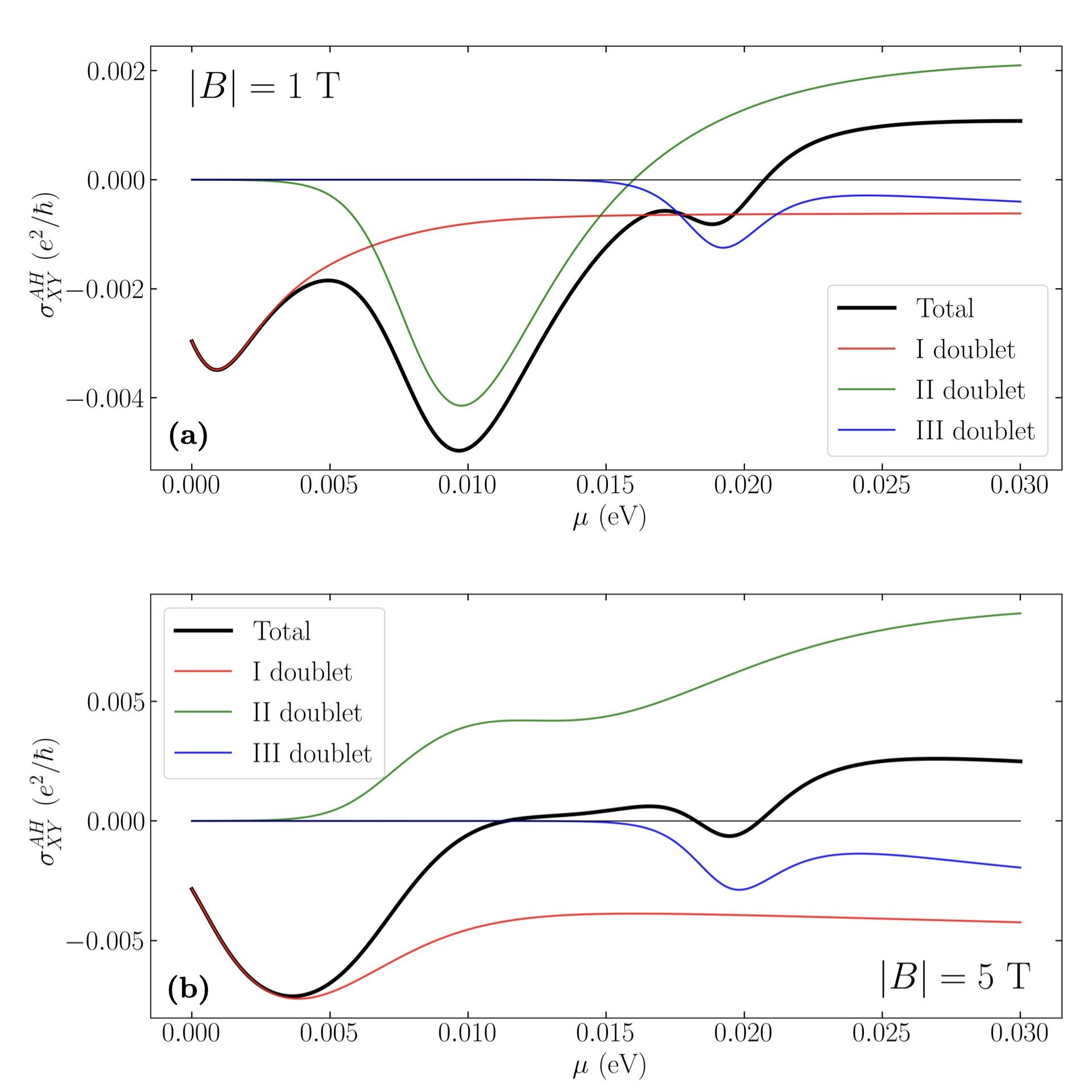}
    \caption{AHC as a function of the chemical potential $\mu$ when $\Vec{B}$ is out-of-plane for (a) $|\Vec{B}|=1$ T and (b) $|\Vec{B}|=5$ T (the temperature is fixed to $T=10$ K). 
    The colored lines represent the contributions from the specific bands.
    }
    \label{cond_anom}
\end{figure}
The magnetic field changes the eigenstates and therefore enters implicitly the Berry curvature. The intrinsic AHE occurs in the moderately dirty regime in which AHC becomes non-dissipative, i.e. independent of the scattering rate~\cite{Tokura2007}. In this regime, the Berry phase of the quasiparticles on the Fermi surface acts as an effective magnetic field generating a transverse momentum. 
In the dirty regime, however, the extrinsic AHC is proportional to the longitudinal conductivity. The extrinsic to intrinsic crossover occurs when the scattering rate becomes comparable to Fermi energy~\cite{al2021oxygen}. In general, insurgence of point defects, such as oxygen vacancies and intermixed cations at LAO/STO interfaces can be controlled e.g. with an amorphous WO$_3$ overlayer to increase the electron mobility~\cite{cavigliamob}. Thus the intrinsic contribution can be made evident.
\\Our main result is shown in Fig.~\ref{contour_inplane} where we plot the AHC as a function of the chemical potential in the presence of an in-plane magnetic field. First, in panel (a) we observe a highly non-linear behavior as a function of the gate voltage. We observe a modulation of the conductance which comes from the sum of the three contributions associated to the doublets of the electronic band structure in Fig.~\ref{fig:band_structure}(d).
In panel (b) we show that this behaviour extends over a wide range of magnetic field magnitudes. In panel (c) we show the periodic behaviour of the AHC by varying the direction of the in-plane magnetic field. The pattern reflects the $C_{3v}$ symmetry of the system; the conductance vanishes when the magnetic field is aligned along the high symmetry lines $\Gamma\rm{K}$($\Gamma\rm{K}^\prime$) due to a residual mirror symmetry which forces the AHC to vanish. The presence of a tetragonal field has been taken into account in appendix~\ref{Appendice-tetra} and in that case these zero conductance lines are reduced since the symmetry itself is reduced. 
The non-linearity and the strong modulation with applied magnetic field stems from two major ingredients: the almost-degenerate multiband structure and the complex topological structure of the Berry curvature.
\\To discuss the relation between the AHC and Berry curvature, we show the AHC $\sigma_{XY}^{\rm{AH}}(\mu)$ as a function of the chemical potential for an out-of-plane magnetic field in Fig.~\ref{cond_anom}. In this case the problem is simplified since the magnetic field is aligned along the (111) direction and thus $J_{111}$ is preserved. The conductivity has a non-linear behavior, as in the case of an in-plane magnetic field, and exhibits more than one dip by varying the chemical potential. Differently from the first dip, the second one is very sensitive to the magnitude of the magnetic field and progressively raises on increasing the magnetic field, until it completely disappears. This can be qualitatively explained by the stronger dependence of the second doublet band splitting on the magnetic field due to its $J=3/2$ character.  This is in contrast with the first and third doublet whose splitting is dominated by the Rashba term. This permits us to conclude that the AHE can be optimized by a gate voltage, at a fixed magnetic field, by moving it close to the second doublet.
\begin{figure}[!t]
    \centering
    \includegraphics[width=0.45\textwidth]{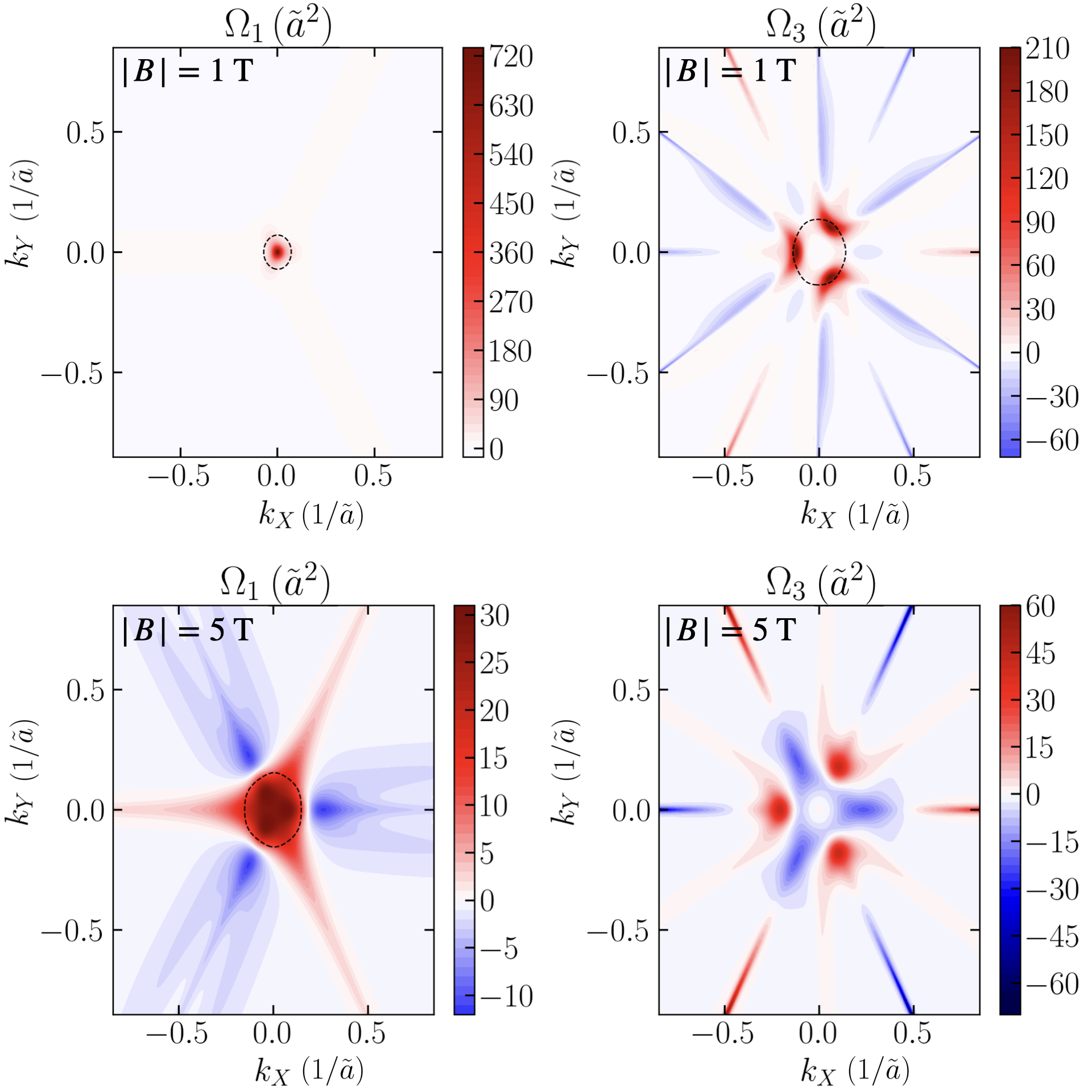}
    \caption{Berry curvatures of the first and the third band at $|\Vec{B}|=1$ T and $|\Vec{B}|=5$ T. The dashed lines correspond to the Fermi contours, such that when they encircle a maximum of the Berry curvature, a dip appears in the Hall conductance.}
    \label{berry_curv}
\end{figure}
\\The dips in the conductivity appears when the Fermi surface encircles one or more peaks of the Berry curvature (see Fig.~\ref{berry_curv}).
The sum of the Berry curvatures of the two bands within the same doublet is typically small, due to partial cancellation (which would be exact under time-reversal symmetry). The splitting of the bands of one doublet can be taken as a small parameter, $\varepsilon_{2\vek}=\varepsilon_{1\vek}+\delta\varepsilon_{\vek}$. Thus the Fermi distribution, namely the Heaviside function at zero temperature, is expanded as $\Theta(\mu-\varepsilon_{1\vek}-\delta\varepsilon_{\vek})\sim \Theta(\mu-\varepsilon_{1\vek})+\delta(\mu-\varepsilon_{1\vek})\delta\varepsilon_{\vek}$. Since $\Omega_2=-\Omega_1+\delta\Omega$, the contribution of a single doublet to the AHC is
\begin{equation}
    \sigma_{XY}^{\text{AH},I}=-\frac{e^2}{\hbar}\int_{BZ}\Omega_{2\vek} \delta(\mu-\varepsilon_{1\vek})\delta\varepsilon_{\vek}+\delta\Omega \Theta(\mu-\varepsilon_{1\vek})\; \frac{d^2\vek}{(2\pi)^2},
\end{equation}
where we can identify the first term as a ring contribution and the second one as an area contribution. The ring contribution is maximum where the curvature is peaked, and leads to the peaks in the AHC; at higher fillings, the area contribution saturates to a constant value and dominates, as can be seen from Fig.~\ref{cond_anom}. An analysis of the two contributions is provided in appendix~\ref{Appendice-contributions}. The change in sign of the conductance comes from the competition of the different doublets contribution: this reflects the importance of the multi-orbital character of the interface for the tunability of the AHE.
\begin{figure}
    \centering
    \includegraphics[width=0.45\textwidth]{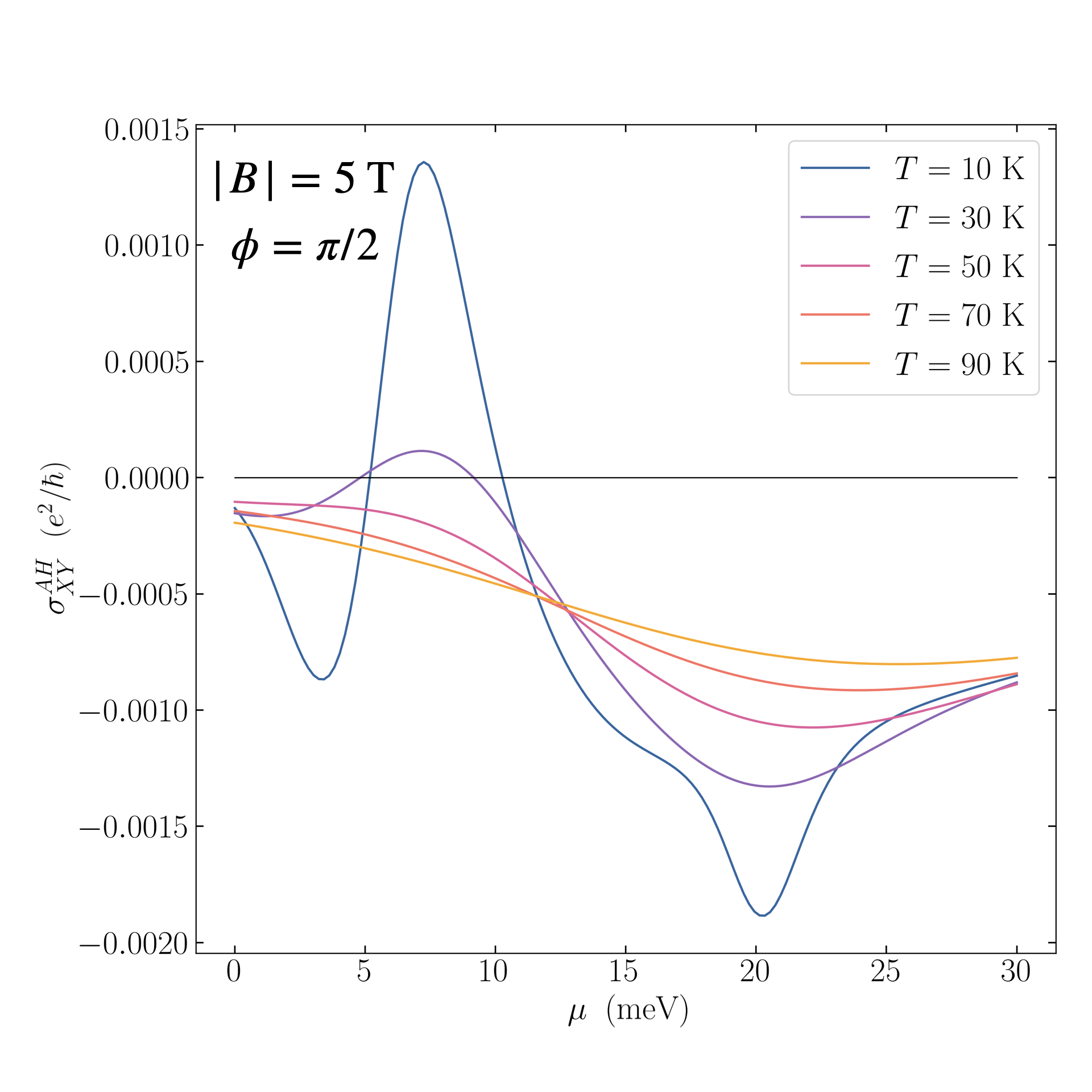}
    \caption{Anomalous Hall conductance as a function of the chemical potential fixing $|\vec{B}|=5$ T and $\phi=\pi/2$ for different values of temperature $T$.}
    \label{temp}
\end{figure}
The out-of-plane magnetic field also gives a conventional Hall conductance which can hinder the anomalous one since its order of magnitude has been verified to be of the order of $ 10^4 e^2/\hbar$ when $|\vec{B}|=5$ T and it cannot be easily subtracted by the total conductance experimentally since it is not linear. This is due to the multiband structure and the presence of the Berry curvature which modifies the element of volume of the phase-space for each band~\cite{2018APS..MARF40002T}. Therefore the major evidence of AHE comes from an in-plane magnetic field, since the results shown in Fig.~\ref{contour_inplane} are not affected by the conventional contribution.
\\Moreover, the AHE strongly depends on the temperature as shown in Fig.~\ref{temp}, in which  we notice the disappearance of the sign change above 30 K. However, this temperature is much higher than the one at which the experiments typically operate. The picture shows that the non-linear behaviour is still present, but the modulation is strongly suppressed by the temperature, since the ring contribution is absent due to the smearing of the Fermi function. We point out that the changing temperature not only affects the Fermi function, but also the mobility of the bands and the scattering rate with impurities. These effects can overcome the intrinsic contribution we evaluated, making more difficult to isolate it at higher temperatures.
The sign change also disappears by including strong Coulomb electron-electron correlations as detailed in appendix~\ref{Appendice-Correlazioni}.
\section{Discussion and conclusions}\label{sec:concl}
We have pointed out and discussed a gate tunable AHE in an oxides interface. The multiband structure of (111) LAO/STO is characterized by a strong near-degeneracy at low fillings and momenta, as well as a strong dependence of the orbital angular momentum on the filling. The combination of these features is responsible for a generalized Rashba coupling in an external electric field. The SOC, which can be quite large in oxides, naturally converts this coupling into a novel Rashba coupling to the total angular momentum. As a first consequence of this Rashba coupling here, we have shown the presence of a spin and angular momentum patterns in- and out-of-plane. Second, we have shown the emergence of an AHC in the presence of an in plane magnetic field due to a non-trivial topology of the Berry curvature.
From the study of the non-linearity of the AHC, one can probe the Berry curvature in the BZ, reconstructing its structure for the different bands.
Finally, the generalized Rashba coupling can strongly affect the behavior of these materials in the superconducting phase in analogy with the case of (001) LAO/STO interface \cite{perroni1,perroni2}.   
\\Our study on LAO/STO interface easily extends to other interfaces. The main requirement is the trigonal geometry and the quenching of the $t_{2g}$ orbitals which is also found in KaTiO$_3$/SrTiO$_3$~\cite{bruno2019band}, as an example. This work opens the way to a general study of the {\it spin-orbitronic} effects induced by the generalized Rashba coupling in oxides interfaces and paves the way to its experimental verification via a gate tunable AHE.
\section*{Acknowledgments} We thank A.Caviglia for insightful suggestions and discussions.
\clearpage
\appendix
\onecolumngrid
\section{Derivation of the Rashba term}\label{Appendice-Electric}
In this section we will derive the effective Rashba term originated from the combination of the SOC and the inversion symmetry breaking due to the electric field.
Here we will first give some details of the calculation of the hopping elements originated from the modification of the orbital shape due to the electric field, and thus the matrix elements of the Hamiltonian~(\ref{electric}).
As we discussed in the main text, we follow the calculation in~\cite{shanavas2014theoretical}, which was referred to another geometry, therefore we will summarize the general idea and give explicitly the results applied to the (111) STO geometry.
\\We need to diagonalize the Hamiltonian $H=H_0+\eta H_E$, where $H_0=H_{\text{TB}}+H_{\text{SO}}+H_{\text{TRI}}$ (expressed in real space), $H_E=-eE_jr_j$ and $\eta=1$ is the small parameter of the perturbation expansion. In our system the electric field can be expressed as $\Vec{E}=E_0(1,1,1)/\sqrt{3}$. It can be considered as a small perturbation causing the atomic orbitals localized at the position $\Vec{R}$ to change in the form $\ket{\phi_{\alpha\Vec{R}}}=\ket{\phi^0_{\alpha\Vec{R}}}+\eta\ket{\phi^I_{\alpha\Vec{R}}}$, where $\ket{\phi^0_{\alpha\Vec{R}}}$ are the unperturbed atomic orbitals.
The hamiltonian matrix to first order in $\eta$ changes in the following way:
\begin{equation}
\begin{split}
    H_{\alpha\beta}=& \bra{\phi^0_{\alpha\Vec{0}}}H_0\ket{\phi^0_{\beta\Vec{R}}}+\eta \bra{\phi_{\alpha\Vec{0}}^I}H_0\ket{\phi_{\beta\Vec{R}}^0}+\\
    &\eta \bra{\phi_{\alpha\Vec{0}}^0}H_0\ket{\phi_{\beta\Vec{R}}^I}+
    \eta\bra{\phi^0_{\alpha\Vec{0}}}H_E\ket{\phi_{\beta\Vec{R}}^0}+O(\eta^2).
    \label{envelop}
\end{split}
\end{equation}
The latest electrostatic term can be neglected since it couples the same orbital to different sites; this is due to the distance between neighbouring sites which causes the matrix elements of $H_E$ to be exponentially suppressed.
\\The modification of the atomic orbitals can be expressed as
\begin{equation}
    \ket{\phi_{\alpha\Vec{R}}^I}=\sum_{\beta\neq\alpha}\frac{\langle \phi_{\beta\Vec{R}}^0|H_E|\phi_{\alpha\Vec{R}}^0\rangle}{\epsilon_\alpha-\epsilon_\beta}\ket{\phi^0_{\beta\Vec{R}}},
    \label{correction eigen}
\end{equation}
evaluated on the same site and where $\alpha$ runs over the $d$ orbitals, while $\beta$ runs over all the other orbitals; here $\epsilon_\alpha$ are the on-site energies of the $\alpha$-th orbital.
We emphasize that the distortion of the orbitals is a completely local, on-site effect.
The angular part of the matrix element in Eq.~(\ref{correction eigen}) can be expressed in terms of the Gaunt Coefficients using the spherical harmonics $Y_m^l$ of the wavefunctions in Eq.~(\ref{correction eigen})
\begin{equation}
    C_{lm,l'm',l''m''}=\int Y_m^l (Y^{l'}_{m'})^*Y_{m''}^{l''} d\Omega
\end{equation}
in which $Y_m^l$ comes from the electric field, $Y_{m''}^{l''}$ comes from $\ket{\phi_{\alpha\Vec{R}}^0}$, which is a $d$ orbital with $l''=2$, and $Y_{m'}^{l'}$ comes from $\ket{\phi_{\beta\Vec{R}}^0}$.
The coefficients vanish unless $l'=l''\pm1$, since $l=1$.
\\The radial part of the matrix element is included in the following coefficient defined as
\begin{equation}
\eta_{\beta}=\frac{|e||\Vec{E}|}{\epsilon_{3d}-\epsilon_\beta}\int R_{3d}(|\Vec{r}|)|\Vec{r}|^3 R_\beta{(|\Vec{r}|)}d|\Vec{r}|.
\end{equation}
where $R_\beta{(|\Vec{r}|)}$ is the radial part of the wavefunction.
\\One can compute the corrections to the $t_{2g}$ states due to the perturbation obtaining~\cite{notaShava}
\begin{equation}
\label{corrections}
    \ket{d_{xy}^I}=\frac{\eta_p}{\sqrt{15}}(\ket{p_x}+\ket{p_y})+\frac{\eta_f}{\sqrt{7}}\left[\frac{2\sqrt{2}}{\sqrt{5}}(\ket{f_{x^2y}}+\ket{f_{xy^2}})+\ket{f_{xyz}}\right].
\end{equation}
and obtaining $\ket{d_{yz}^I}$ and $\ket{d_{zx}^I}$ with cyclic permutations of $x$, $y$ and $z$.
It is worthwhile noticing that the $\eta_f$ is small compared to $\eta_p$~\cite{shanavas2014theoretical}, so that in the following we will consider only the $\ket{p_i}$ corrections.
\\Now we can calculate the matrix elements $(\bra{\phi_{\alpha\Vec{0}}^I}H_0\ket{\phi_{\beta\Vec{R}}^0}+\bra{\phi_{\alpha\Vec{0}}^0}H_0\ket{\phi_{\beta\Vec{R}}^I})$. 
Let us assume for the $H_0$ eigenstates the following notation in the expansion of the atomic orbitals expressed as a function of the quasi-momentum $\Vec{K}$:
\begin{equation}
    \ket{\phi_{i\Vec{K}}^{\text{Ti}_\alpha}}=\ket{i_{\Vec{K}},\alpha}=\frac{1}{\sqrt{N}}\sum_{\Vec{r}}e^{-i\Vec{K}\cdot \Vec{r}}\ket{i(\Vec{r}-\Vec{\delta}_\alpha)},
\end{equation}
where we emphasized the fact that two non-equivalent Ti atoms in the same crystal cell are separated by a base vector, so that $\Vec{\delta}_\alpha=\{(0,0),(0,-a_0)\}=\{\Vec{0},\Vec{\delta}\}$ in the 2D space.
To shorten the notation, we write the corrections of the $d$ orbitals defined in Eq.~(\ref{corrections}) as
\begin{equation}
    \ket{d_{i}(\Vec{r})^I}=\mathcal{A}_{ij}\ket{p_{j}(\Vec{r})}.
\end{equation}
First of all, let us calculate these matrix elements over the nearest neighbour atoms, coupling the Ti atoms from different layers.
Therefore we need to compute the following matrix elements
\begin{dmath}
    \bra{d_{i,\Vec{K}}^I,1}H_0\ket{d_{j,\Vec{K}},2}+ \bra{d_{i,\Vec{K}},1}H_0\ket{d_{j,\Vec{K}}^I,2}=
    \sum_{\Vec{a}}\mathcal{A}_{ik}^*\mel{p_k(\Vec{r})}{H_0}{d_j(\Vec{r}-\Vec{a}-\Vec{\delta})}e^{-i\Vec{K}\cdot\Vec{a}}
    +\mathcal{A}_{jk}\mel{d_i(\Vec{r})}{H_0}{p_k(\Vec{r}-\Vec{a}-\Vec{\delta})}e^{-i\Vec{K}\cdot\Vec{a}},
    \label{KTiKti}
\end{dmath}
where $\Vec{a}$ are the vectors connecting the nearest neighbour elementary cells.
For every $\Vec{a}$ we can associate a certain direction in terms of the 2D-dimensionless quasi-momentum $\vek=(k_X,k_Y)$ via projection of the (111) plane:
\begin{itemize}
    \item for $\Vec{a}=\Vec{a}_x=a_0(1,0,0)\rightarrow \Vec{K}\cdot \Vec{a}_x=-\frac{\sqrt{3}}{2}k_X-\frac{3}{2}k_Y$;
    \item for $\Vec{a}=\Vec{a}_y=a_0(0,1,0)\rightarrow \Vec{K}\cdot \Vec{a}_y=\frac{\sqrt{3}}{2}k_X-\frac{3}{2}k_Y$;
    \item for $\Vec{a}=\Vec{a}_z=a_0(0,0,1)\rightarrow \Vec{K}\cdot \Vec{a}_z=0$;
\end{itemize}
where the latter is zero due to the fact that the two atoms connected by $\Vec{a}_z$ are within the same elementary cell.
\\We can evaluate the matrix elements appearing in Eq.~(\ref{KTiKti}) for the $t_{2g}$ orbitals and thus obtain the following matrix elements connecting Ti$_1$ with Ti$_2$
\begin{dmath}
    \mel{i_{\vek},\mathrm{Ti}_1}{H_0}{j_{\vek},\mathrm{Ti}_2}=
    \eta_p\frac{V_{pd\pi}(\sqrt{2})^{7/2}}{\sqrt{15}}
    \begin{pmatrix}
    0 & -2i e^{i\frac{3}{2}k_Y} \sin{(\frac{\sqrt{3}}{2}k_X)}& 1-e^{\frac{i}{2}\left(\sqrt{3}k_X+3k_Y\right)} \\
    2i e^{i3/2k_Y} \sin{(\frac{\sqrt{3}}{2}k_X)} & 0 & 1-e^{-\frac{i}{2}\left(\sqrt{3}k_X-3k_Y\right)}\\
    -1+e^{\frac{i}{2}\left(\sqrt{3}k_X+3k_Y\right)}  & -1+e^{-\frac{i}{2}\left(\sqrt{3}k_X-3k_Y\right)} & 0\\
    \end{pmatrix},
    \label{t1hopt2}
\end{dmath}
where we expressed the matrix using the basis $\{\ket{d_{yz}},\ket{d_{zx}},\ket{d_{xy}}\}$ and introduced the Slater-Koster (SK) integral $V_{pd\pi}$. 
Let us now compute the next-to-nearest-neighbour terms (NNN), coupling the Ti atoms belonging to the same layer. In this case the direction which connects two different atoms, expressed in $\vek$ space, are:
\begin{itemize}
    \item for $\Vec{a}=\Vec{a}_1=a_0(0,+1,-1)\rightarrow \kappa_1=-\frac{\sqrt{3}}{2}k_X+\frac{3}{2}k_Y$
    \item for $\Vec{a}=\Vec{a}_2=a_0(-1,0,+1)\rightarrow \kappa_2=-\frac{\sqrt{3}}{2}k_X-\frac{3}{2}k_Y$;
    \item for $\Vec{a}=\Vec{a}_3=a_0(+1,-1,0)\rightarrow \kappa_3=\sqrt{3}k_X$.
\end{itemize}
A straightforward calculation leads to two different matrices $H_{E\pi}$ and $H_{E\sigma}$ which are regulated by the two different SK parameters $V_{pd\pi}$ and $V_{pd\sigma}$ for the NNN contribution:
\begin{dmath}\small
    H_{E\pi}=\eta_p\frac{2i}{\sqrt{15}}V_{pd\pi}
    \begin{pmatrix}
        0 & -(\sin(\kappa_1)+\sin(\kappa_2)+2\sin(\kappa_3)) & (\sin(\kappa_1)+2\sin(\kappa_2)+\sin(\kappa_3))\\
        (\sin(\kappa_1)+\sin(\kappa_2)+2\sin(\kappa_3)) & 0 & -(2\sin(\kappa_1)+\sin(\kappa_2)+\sin(\kappa_3))\\
        (\sin(\kappa_1)+2\sin(\kappa_2)+\sin(\kappa_3)) & (2\sin(\kappa_1)+\sin(\kappa_2)+\sin(\kappa_3)) & 0
        \end{pmatrix}\normalsize
    \label{electric_pi}
\end{dmath}
\begin{dmath}
    H_{E\sigma}=\eta_p\frac{2i}{\sqrt{15}}\sqrt{3}V_{pd\sigma}
    \begin{pmatrix}
        0 & (\sin(\kappa_1)+\sin(\kappa_2)) & -(\sin(\kappa_1)+\sin(\kappa_3))\\
        -(\sin(\kappa_1)+\sin(\kappa_2)) & 0 & (\sin(\kappa_2)+\sin(\kappa_3))\\
        (\sin(\kappa_1)+\sin(\kappa_3)) & -(\sin(\kappa_2)+\sin(\kappa_3)) & 0
        \end{pmatrix},
    \label{popo}
\end{dmath}
where we have neglected the spin and the layer degree of freedom, since it is diagonal in these labels. 
Referring to the approximation made by Ref.~\cite{shanavas2014theoretical}, we can use the form for the SK parameters 
\begin{equation}
    V_{pd\pi/\sigma}=n_{pd\pi/\sigma}\frac{\hbar^2r_a^{3/2}}{m|a|^{7/2}}
\end{equation}
where $r_a=10.8$ nm, $\hbar^2/m=7.62\times 10^{-2}$ eV nm$^2$, $n_{pd\sigma}=-3.14$ and $n_{pd\pi}=1.36$ and $|a|$ is the distance between the two atoms. Thus $V_{pd\pi}=0.028$ eV and $V_{pd\sigma}=-0.065$ eV.
Furthermore since the magnitude of the electric field can be expressed via the electric potential between the two layers $v$ as $|\Vec{E}|=v\sqrt{3}/a_0$; we extract the value of $\eta$ from the parameters as $\eta_p\sim \frac{|\Vec{E}|}{10\text{ eV/nm}}\sim0.09$ since $v=0.2$ eV and $a_0=3.905$ nm.
\subsection*{Expansion for small fillings.}
The previous expression is the full inversion symmetry breaking Hamiltonian, but it does not have the more common form of the Rashba interaction. We will perform another perturbation calculation in order to derive the usual expression of the Rashba term. We can evaluate exactly the eigenvectors of the first two dominant terms in the complete Hamiltonian which are the TB Hamiltonian and the electrostatic contribution due to the local $\Vec{E}$ action on the states, which is denoted as $H_{\text{TB}}+H_{v0}$. 
\\The whole matrix $H_{\text{TB}}+H_{v0}$, which is $12\times12$, admits as eigenstates
\begin{equation}
    \ket{\psi_{i\sigma,\vek}{\pm}}=\alpha_i(\vek)e^{i\phi_i(\Vec{k})}\ket{d_{i1\sigma,k}}+ \beta_i^{\pm}(\vek) \ket{d_{i2\sigma,k}}
    \label{vectors66},
\end{equation}
with
\begin{align}\label{coefficients66}
      \nonumber&\alpha_i^{\pm}(\vek)=\frac{|\epsilon_i(\vek)|}{\sqrt{2|\epsilon_i(\vek)|^2+\frac{v^2}{2}\pm v\sqrt{\frac{v^2}{4}+|\epsilon_i(\vek)|^2}}};
      \\
      &\beta_i^{\pm}(\vek)=\frac{\left(\frac{v}{2}\pm\sqrt{\frac{v^2}{4}+|\epsilon_i(\vek)|^2}\right)}{\sqrt{2|\epsilon_i(\vek)|^2+\frac{v^2}{2}\pm v\sqrt{\frac{v^2}{4}+|\epsilon_i(\vek)|^2}}};
      \\
      \nonumber&\phi_i(\Vec{k})=\arg[\epsilon_i(\vek)]
\end{align}
where the orbitals are labeled by the index $i$ and the spin using the index $\sigma$.
The corresponding eigenvalues are
\begin{equation}
    \mathcal{E}_i^{\pm}(\vek)=\pm\sqrt{\frac{v^2}{4}+|\epsilon_i(\vek)|^2}.
    \label{eigenval66}
\end{equation}
These states are well separated in six upper bands $\ket{\psi_{i\sigma,k}^{+}}$ and six lower bands $\ket{\psi_{i\sigma,k}^{-}}$. From now on we will take into account only the lower states, so we will neglect the $\pm$ label. 
The six lower bands are degenerate at the origin. Consequently, for sufficiently small values of $\Vec{k}$, the TB hamiltonian splits the bands only by an amount of the order of $t_3 |\Vec{k}|^2$. 
\\In order to obtain the Electric field Hamiltonian on the six lower bands for low fillings, we simultaneously linearize the Electric field Hamiltonians~(\ref{electric_pi}) and~(\ref{popo}) as a function of $\vek$ and evaluate its matrix elements among the six lower states in Eq.~(\ref{vectors66}), evaluated for $\vek=0$. The result is the following linear Hamiltonian:
\begin{equation}
\begin{split}
\centering
    &(H_E)_{ij}=-i\mathcal{F}\varepsilon_{ijk}\kappa_k, \quad \text{where}\\
    &\mathcal{F}=\frac{2\eta_p}{\sqrt{15}}\left(V_{pd\pi}(1+2^{7/8}\alpha\beta\cos{(\phi}))+\sqrt{3}V_{pd\sigma}\right),
\end{split}
\end{equation}
$\Vec{\kappa}=(\kappa_1,\kappa_2,\kappa_3)$ as defined above and $\alpha$, $\beta$ and $\phi$ are the Eqs.~(\ref{coefficients66}) evaluated for $\vek=0$.
\\Identifying now the matrix elements of the orbital angular momentum $\hat{L}$ matrices, we can rewrite this term as:
\begin{equation}
    H_{E}=\frac{3}{\sqrt{2}}\mathcal{F} (\vek\times\hat{L})\cdot\frac{\Vec{E}}{|\Vec{E}|}\otimes\mathbb{1}_{\sigma\sigma'}.
    \label{eq_kl}
\end{equation}
\begin{figure}
    \centering
    \includegraphics[width=0.50\textwidth]{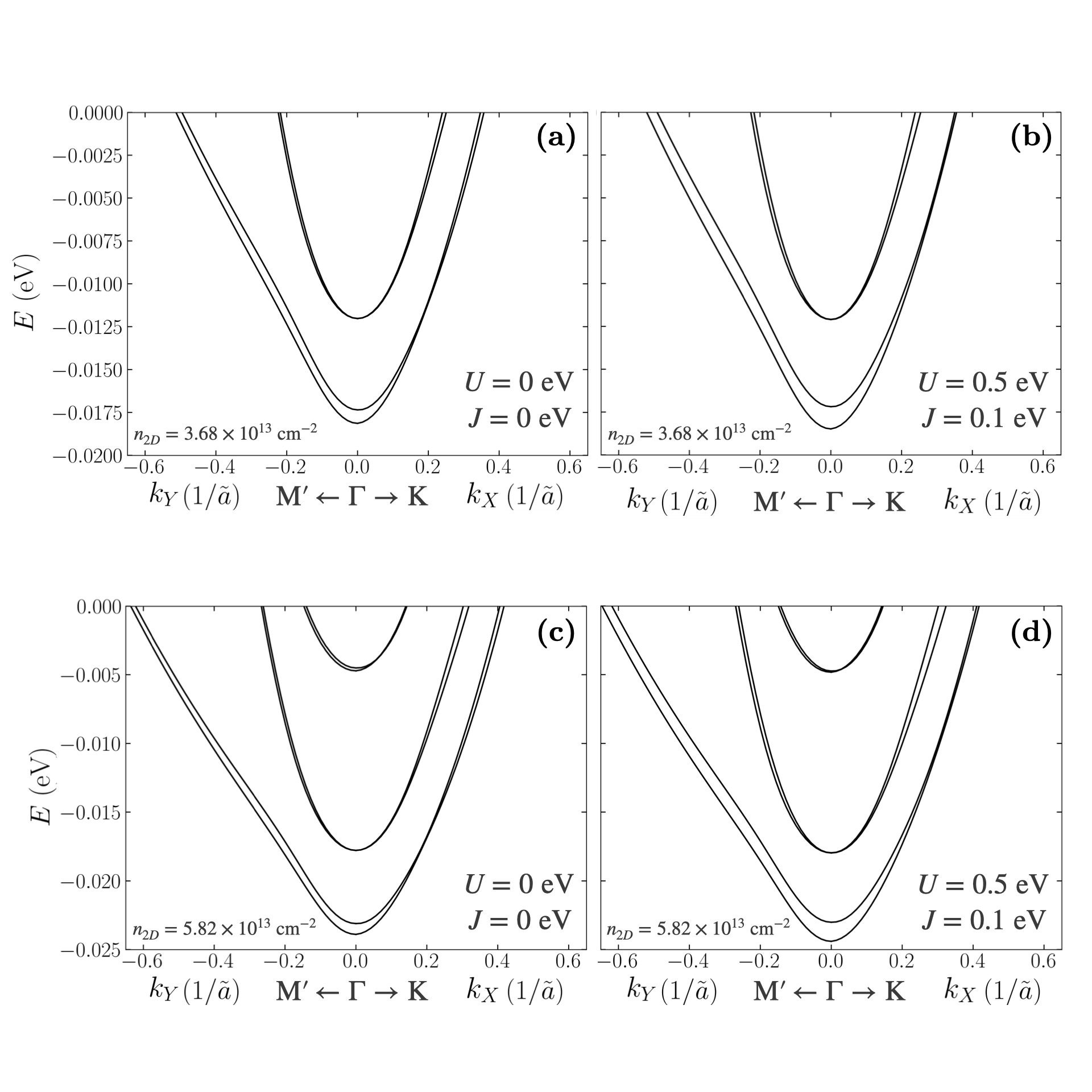}
    \caption{Electronic band structure evaluated for (a-c) $U=0$ eV and $J=0$ eV and (b-d) $U=0.5$ eV and $J=0.1$ eV with $B=5$T and $\phi=\pi/2$. $E=0$ corresponds to the Fermi level when the 2D electron density is (a-b) $n_{2D}=3.68\times10^{13}$ cm$^{-2}$ or (c-d) $n_{2D}=5.82\times10^{13}$ cm$^{-2}$.}
    \label{bandsRenormalized}
\end{figure}
This term, the so-called orbital Rashba coupling~\cite{revieworbitronic2021}, has the same structure of the Rashba coupling, but it involves $\hat{L}$ instead of $\hat{S}$. The same coupling has been derived using another electrostatic argument~\cite{park2011orbital,kim2013microscopic,kim2014nature,hong2015quantitative}.
Having introduced the notation of the angular momentum we can write also the reduced $6\times6$ matrix of the TB over the states~(\ref{vectors66}) using the same notation:
\begin{equation}
    H_{\text{TB}}=\sum_{i}\mathcal{E}_i (\mathbb{1}-L_i^2)\otimes\mathbb{1}_{\sigma\sigma'},
    \label{TBintermsofL}
\end{equation}
where $\mathcal{E}_i$ are the quadratic expansion of Eq.~(\ref{eigenval66}). Also $H_{\text{TRI}}$ admits an expression of the form
\begin{equation}
    H_{\text{TRI}}=\Delta(\scalebox{1.25}{$\mathbb{1}$}-\frac{3}{2}L_{111}^2).
    \label{eq_app_delta}
\end{equation}
Now, it is clear that, when $\vek$ is small enough that SOC is the dominant contribution, the previous couplings can be evaluated over the SOC eigenstates, and thus the total angular momentum appears leading to the Hamiltonian
\begin{equation}
    H_{\rm{eff}}(|\vek|\sim0)=-\frac{\Delta}{2} J_{111}^2+\frac{2\mathcal{F}}{3}(\vek\times\hat{J})+\sum_{i}\mathcal{E}_i(\vek) (\nu\mathbb{1}-\tau J_i^2),
\end{equation}
for which $\tau=1/3$ and $\nu=3/4$ when we evaluate this operators on the lower quadruplet states ($J=3/2$, $L=1$ and $S=1/2$).

\section{Role of correlations}\label{Appendice-Correlazioni}
\begin{figure}
    \centering
    \includegraphics[width=0.47\textwidth]{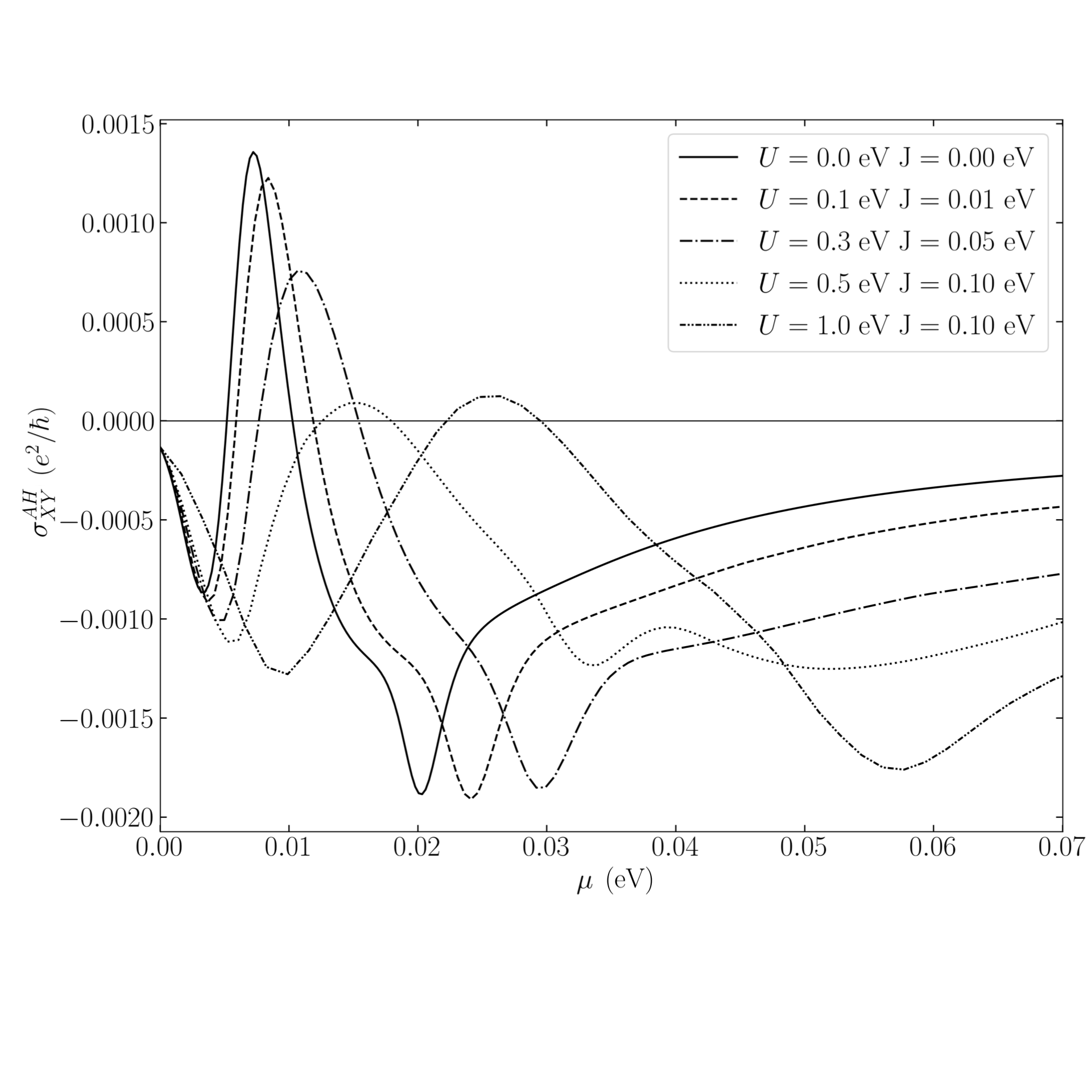}
    \caption{Anomalous Hall conductance as a function of the chemical potential fixing $|\vec{B}|=5$ T and $\phi=\pi/2$ for different values of correlation parameters.}
    \label{correlation_condu}
\end{figure}
The Hubbard Hamiltonian we take into account, written in real space, is the following
\begin{equation}
    H_{\rm{H}}=U\sum_{r,\alpha}n_{r\alpha\uparrow}n_{r\alpha\downarrow}+\frac{1}{2}\sum_{r,\alpha\neq\beta,\sigma\sigma^{\prime}}(U^{\prime}-J\delta_{\sigma\sigma^{\prime}})n_{r\alpha\sigma}n_{r\beta\sigma^{\prime}},
\end{equation}
in which $n_{r\alpha\sigma}$ is the number operator of the state located in the site $r$ occupying the orbital $\alpha$ with spin $\sigma$: we choose the basis spin states to be aligned (or antialigned) with the direction of the magnetic field. This choice is physically motivated by the observation that a spin imbalance in the Hamiltonian can only be generated by the magnetic field. We suppressed the label of the Ti atom due the local nature of the interaction (in other terms, $r$ runs in the same time over the location of the elementary cell and the type of atom). Considering the Hamiltonian in the mean field approximation we obtain, suppressing the index of the site $r$,
\begin{equation}\footnotesize
    H_{\rm{H}}=\sum_{\alpha}n_{\alpha\uparrow}\left(U\nmean{\alpha\downarrow}+\sum_{\beta\neq\alpha}\left(U^{\prime}-J)\nmean{\beta\uparrow}+U^{\prime}\nmean{\beta\downarrow}\right)\right)+(\uparrow\leftrightarrow\downarrow).
\end{equation}
\begin{figure}
    \centering
    \includegraphics[width=0.45\textwidth]{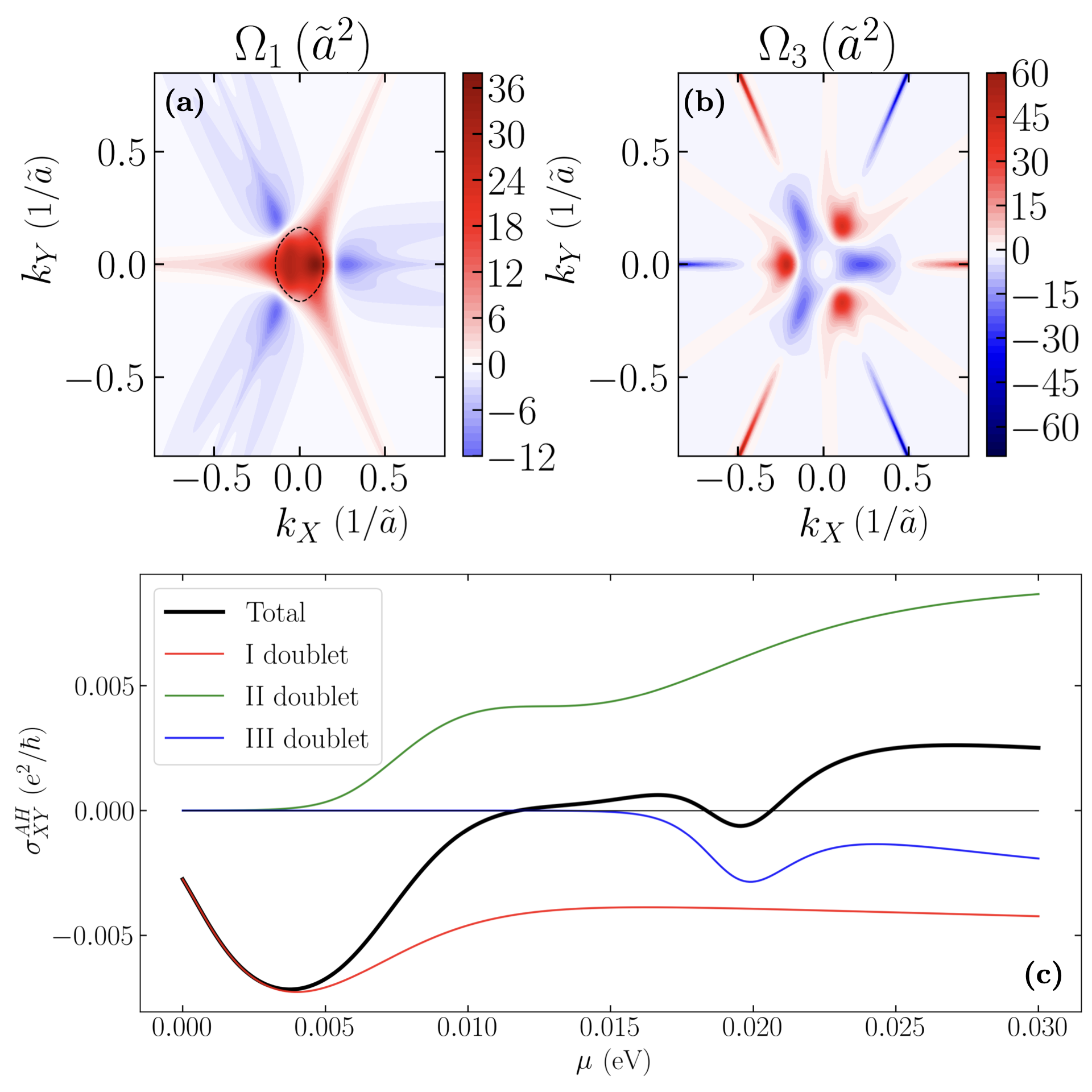}
    \caption{Berry curvatures for (a) the first and (b) the third band when the magnetic field is out-of-plane at $|\Vec{B}|=5$ T and $\zeta=-0.003$ eV. (c) Anomalous Hall conductance as a function of the chemical potential. The dashed line in (a) represents the Fermi surface at the dip of the conductance.}
    \label{Berry_cond_zeta}
\end{figure} 
Due to translational invariance $H_{\rm{H}}$ is local and can be written in $\vek$ space.
By choosing $U^\prime=U-2J$ by rotational invariance~\cite{georges2013strong}, the Hamiltonian has only two free parameters. 
The band structure depends on the fillings, which are evaluated self-consistently. Our algorithm proceeds in the following way: in the absence of correlations we evaluate, for a fixed chemical potential, the electron occupations for each orbital $\langle n^0_{\alpha}\rangle$ and the total occupation $N_0=\sum_{\alpha}\langle n^0_{\alpha}\rangle$. We include the correlations and we evaluate the bands using the $\langle n^0_{\alpha}\rangle$. From this band structure we find the chemical potential at which the total density is equal to $N_0$. We use $\langle n^0_{\alpha}\rangle$ corresponding to this renormalized chemical potential as a starting point for a routine which self-consistently evaluates the correct filling fractions. The renormalized bands in the regime of small fillings considered in the paper are depicted in Fig.~\ref{bandsRenormalized}.
The predicted AH conductivity for different choices of parameters is depicted in Fig.~\ref{correlation_condu}. The correlations do not change the non-linear behaviour of the conductivity, and their main effect is a renormalization of the magnetic field $\vec{B}$ and the electric field $v$ depending on the filling. When we reach the regime of $U=0.5$ eV the behaviour of the third doublet changes showing a new dip and the curve does not change its sign anymore. This behaviour traces a new regime in the correlations which can significantly modify the conductance: this property could be exploited to test the strength of correlations in the (111) LAO/STO interface, a question which is still debated in literature. We notice that the values of U considered in this work are not large, but they are comparable with the electron bandwidth. Furthermore, in LAO/STO systems analyzed in this work, the typical particle densities are low compared to the half-filling regime. Accordingly, the net contribution to the total energy from Hubbard interaction is small. Furthermore, in these density regimes, the effects from polaron dynamics can be important on the electronic states~\cite{perronipol} giving rise to a net lowering of the Hubbard interaction for the electronic system.

\section{Breaking of the $C_{3v}$ symmetry: tetragonal strain}\label{Appendice-tetra}
\begin{figure}
    \centering
    \includegraphics[width=0.45\textwidth]{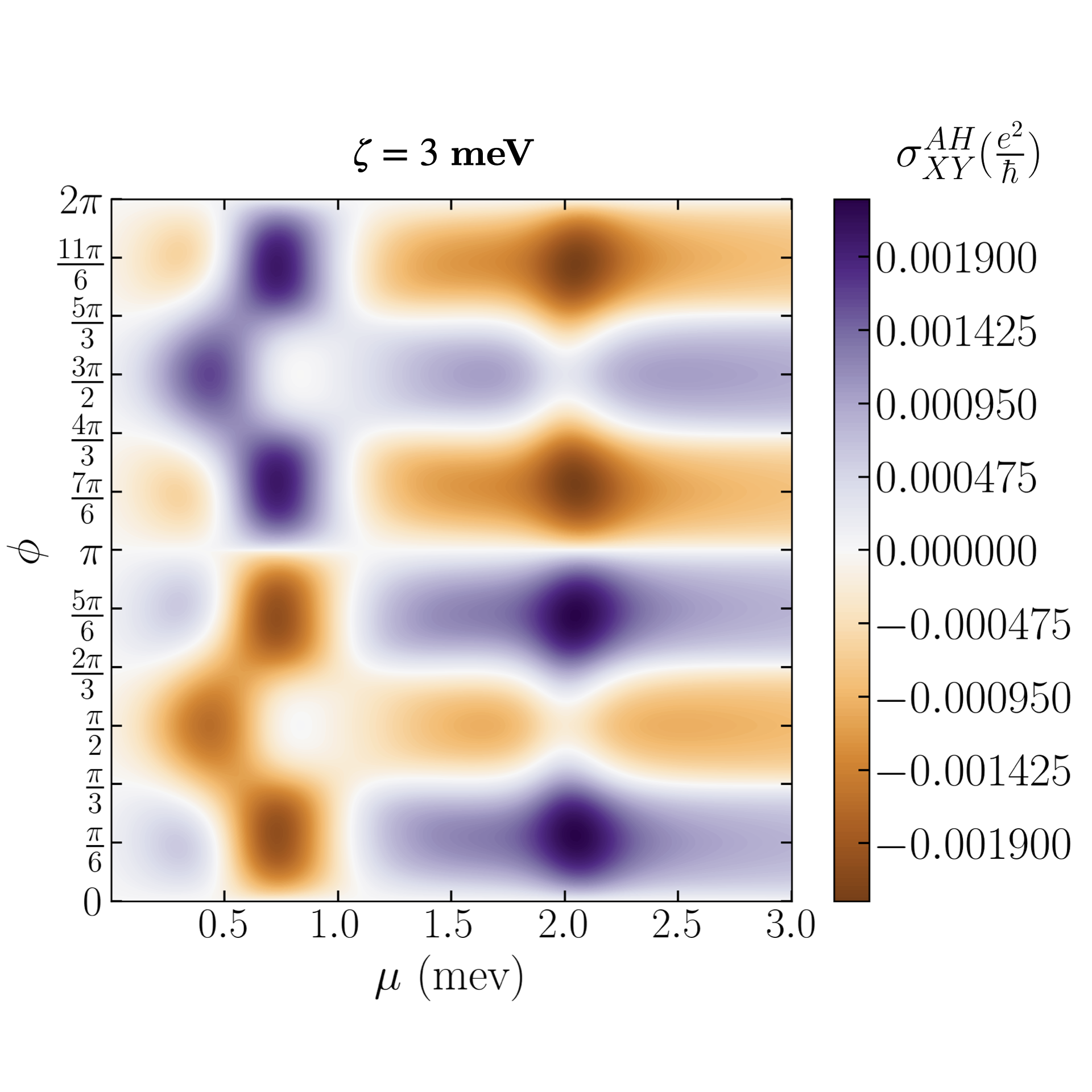}
    \caption{Anomalous Hall conductance as a function of $\mu$ for different values of $\phi$ at $|\Vec{B}|=5$ T, $\zeta=-0.003$ eV and $T=10$ K. The pattern is periodic with a change in sign of the conductance for $\phi=n\pi/2$, with $n\in\mathds{Z}$.}
    \label{vabb}
\end{figure}
\begin{figure*}
    \centering
    \includegraphics[width=0.85\textwidth]{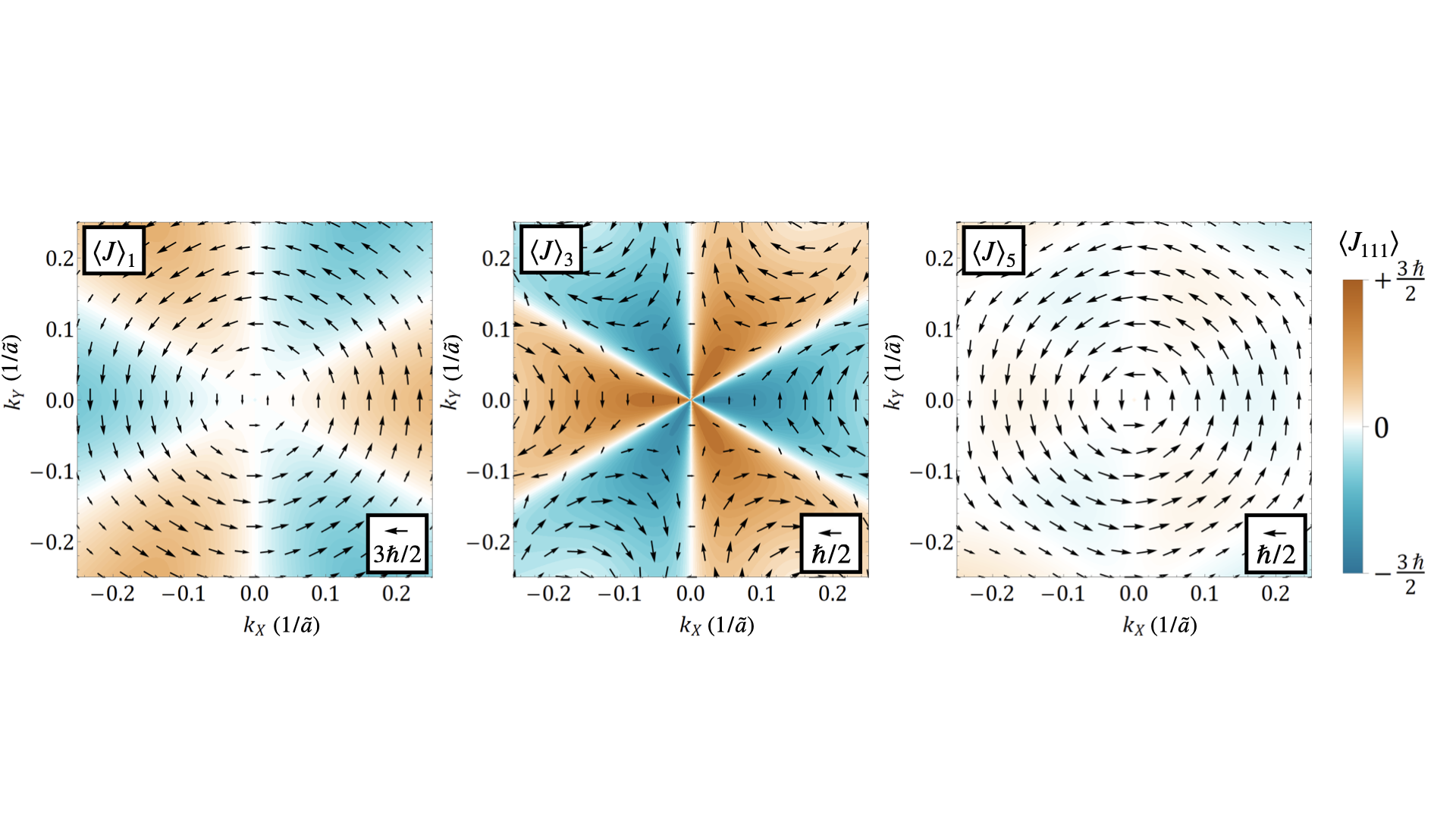}
    \caption{In- and out-of-plane spin and angular momentum modulation for the first, the third and the fifth band. The second, the fourth and the sixth are specular to the ones shown. The in-plane patterns are represented by the arrows and are obtained by computing the mean value of the spin components $J_{[\overline{1}10]}$ and the $J_{[\overline{11}2]}$ over the eigenstates of the chosen band. The colours indicate the modulation out-of-plane.}
    \label{J_pattern}
\end{figure*} 
Fig.~\ref{contour_inplane} manifests the $C_{3v}$ symmetry of the system. In figure we observe two different types of white line on which the conductance vanishes. The vertical lines represent the vanishing of the conductance for a fixed chemical potential, by varying the direction of the magnetic field. The zero conductance, for this value of the chemical potential, does not depend on the direction of $\Vec{B}$ since the shape of the Berry curvature is only rotated, while the energy splitting between the bands depends only on the magnitude of $\vec{B}$. The horizontal lines, instead, depends on an accidental residual degeneracy of the system and therefore the Berry curvature cannot be defined along these lines, leading to the vanishing of the conductance. The degeneracy we refer to is the reflection symmetry around $\vec{B}$ on the $\Gamma \rm{M}$($\Gamma\rm{M}^\prime$) line when $\vec{B}$ is aligned to the $\Gamma\rm{K}$($\Gamma\rm{K}^\prime$) line. This effect depends crucially on the $C_{3v}$ symmetry of the system~\cite{notaMirror}. 
\\However, at low temperatures it is possible to have an additional tetragonal distortion which can be comparable in order of magnitude with the other couplings. This induces a tetragonal strain at the interface, resulting in the energetic preference of one orbital, e.g. the $d_{xy}$, with respect to the others. This term will reduce the symmetry of the lattice, changing the prediction we made. In order to take into account this distortion in the system one has to include the following Hamiltonian
\begin{equation}
    H_{z}=\frac{\zeta}{2}\sum_{\vek}\sum_{\alpha,\sigma}d_{xy\hspace{0.05cm}\alpha\sigma,\vek}^{\dagger}d_{xy\hspace{0.05cm}\alpha\sigma,\vek},
\end{equation}
where we choose $\zeta=-0.003$ eV.
This term can be written with the angular momentum notation we used for the derivation of Eqs.~(\ref{eq_kl}), (\ref{TBintermsofL}) and (\ref{eq_app_delta}) by the same procedure
\begin{equation}
    H_z=\frac{\zeta}{2}\left(\scalebox{1.25}{$\mathbb{1}$}-L_{z}^2\right).
    \label{eq:zeta}
\end{equation}
Eq.~(\ref{eq:zeta}) can be further expressed in terms of the total angular momentum $\Vec{J}$ when the SOC is the dominant interaction, obtaining
\begin{equation}
    H_z\sim\frac{\zeta}{3}J_z^2+\text{cost}.
\end{equation}
Since both $J_z^2$ and $J_{111}^2$ appear in the Hamiltonian, it is impossible to select a set of eigenstates which are also eigenstates of one of the component of $\Vec{J}$.
\\When the out-of-plane magnetic field is included into the system the tetragonal strain modifies the shape of the Berry curvature, reducing the symmetry of the system from $C_{3v}$ to a in-plane reflection, as one can see in Fig~\ref{Berry_cond_zeta}(a).
However, this has however limited consequences on the AHC for out-of-plane magnetic field, as one can observe in Fig.~\ref{Berry_cond_zeta}(c). 
\\The most interesting effect is observed for configurations with in-plane magnetic field. The tetragonal strain generally removes the horizontal white lines in Fig.~\ref{vabb}(a) except when the magnetic field is orthogonal to the direction of the in-plane projection of the strain direction. This behaviour is exemplified in Fig.~\ref{vabb}(b) where $\sigma_{XY}^{AH}=0$ only when $\phi=0$ or $\pi$ due to this symmetry protection.

\section{Total angular momentum pattern}\label{Appendice-totAng}
In this section we report the in- and out-of-plane pattern of the total angular momentum $\vec{J}$ for sake of completeness. The mean value of $|\vec{J}|$ saturates to its highest value for the first two doublets to $3/2\hbar$ while for the third one to $1/2\hbar$: which is another sign of the total angular momentum interpretation for the low energy doublets.
\begin{figure*}
    \centering
    \includegraphics[width=0.47\textwidth]{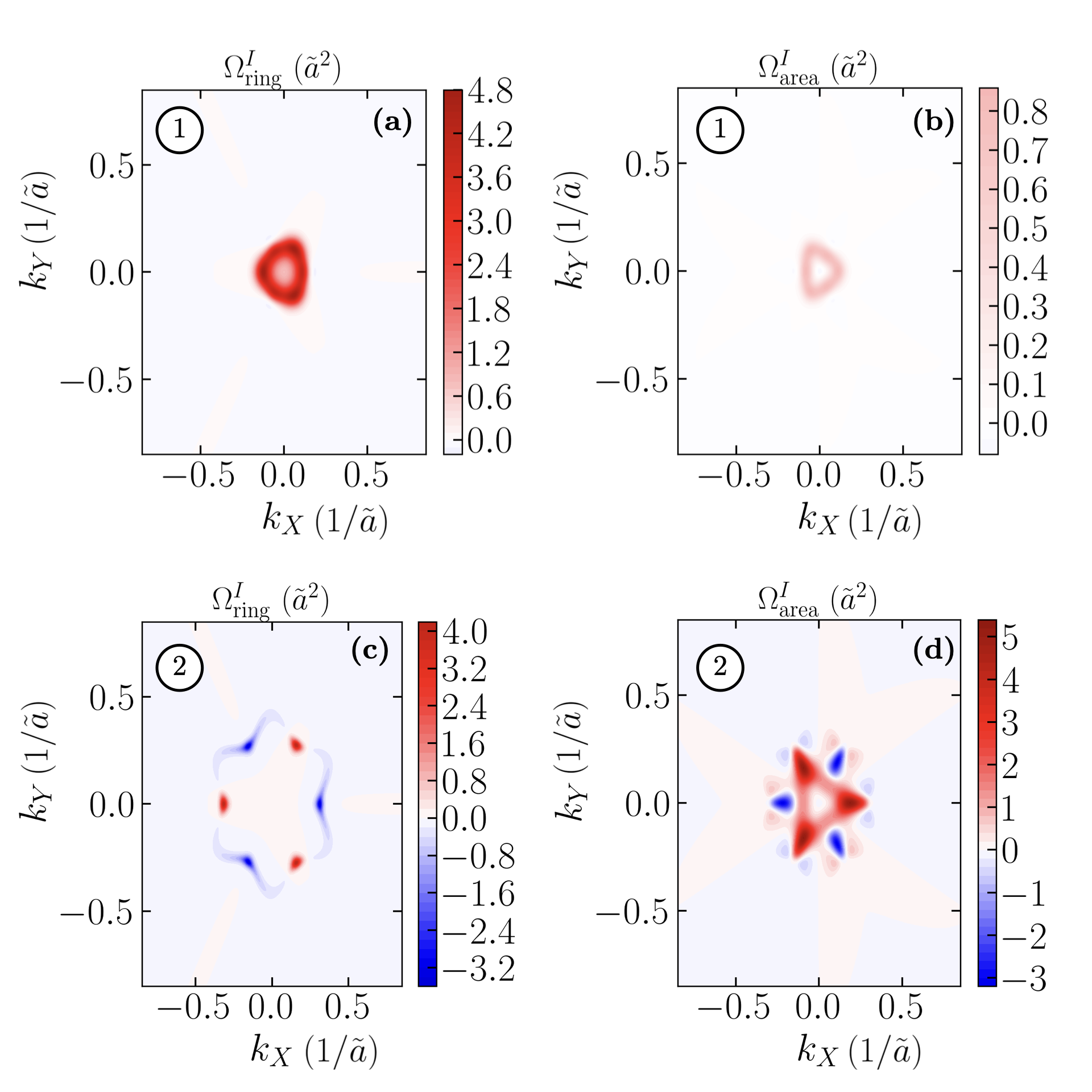}\hfill
    \includegraphics[width=0.47\textwidth]{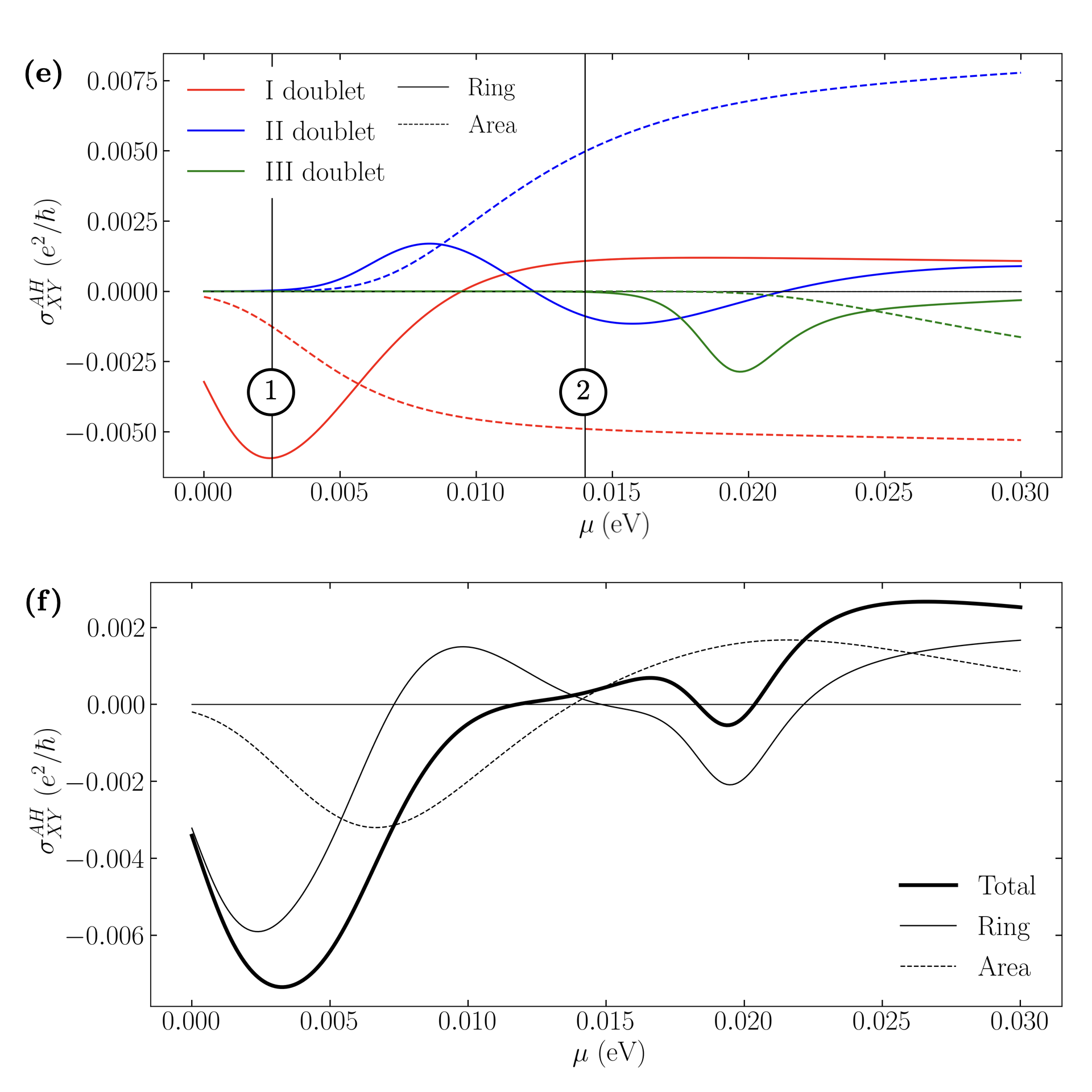}
    \caption{(a-d) Berry-ring and Berry-area of the first doublet for values of the chemical potentials highlighted by the lines in the right panel evaluated for an out-of-plane magnetic field of $|\vec{B}|=5$ T. (e-f) Integrated contribution to the Hall conductivity of the three doublets (red, blue, and green respectively): we show the total contribution from all the doublets in black.}
    \label{fig:contours}
\end{figure*}

\section{Ring and area contribution to the anomalous conductance}\label{Appendice-contributions}
In this section we discuss the different behaviour of the two contributions in the expression
\begin{equation}\footnotesize
        \sigma_{XY}^{\text{AH},I}=-\frac{e^2}{\hbar}\int_{BZ} \left[ \Omega_{2\vek} \delta(\mu-\varepsilon_{1\vek})\delta\varepsilon_{\vek}+\delta\Omega \Theta(\mu-\varepsilon_{1\vek})\right] \frac{d^2\vek}{(2\pi)^2},
    \end{equation}
which represents the AHC due to the first energy doublet. We refer to the first member of the integrand as Berry-ring ($\Omega_{\rm{ring}}^I=\Omega_{2\vek} \delta(\mu-\varepsilon_{1\vek})\delta\varepsilon_{\vek}$) and to the second as Berry-area ($\Omega_{\rm{area}}^I=\delta\Omega \Theta(\mu-\varepsilon_{1\vek})$), and we refer to their integrals as ring and area contribution respectively. 
We extend the previous expression to finite but small temperature to regularize the integral with the delta function:
\begin{dmath}
        \sigma_{XY}^{\text{AH},I}=-\frac{e^2}{\hbar}\int_{BZ} \Omega_{2\vek} \frac{\partial f_{th}}{\partial \varepsilon_{1\vek}}(\varepsilon_{1\vek}-\mu)\delta\varepsilon_{\vek}+
        \delta\Omega f_{th}(\varepsilon_{1\vek}-\mu) \; \frac{d^2\vek}{(2\pi)^2}.
\end{dmath} 
The plots of the Berry-ring and the Berry-area for the first doublet are represented in Fig.\ref{fig:contours}(a-d) for two different $\mu$ corresponding to the energies in right panel. When the chemical potential is fixed to the first value, e.g. the solid red dip in Fig.~\ref{fig:contours}, the Berry-ring contribution is higher in value and is everywhere positive. 
\\The Berry-area contribution is smaller in value, since, the Berry curvature of the two components of the doublet nearly cancel out. Far from the dip in the AHC, the ring contribution saturates to a small value while the area contribution dominates. 
\\The AHC separated into the different contributions for all the three doublets is shown in Fig.~\ref{fig:contours}(e-f).

\bibliographystyle{unsrt}

\begin{thebibliography}{10}
\bibitem{notaJ}
{ The reason is well explained by the effective model in Eq.~(\ref{jmodel}) since
  $\vek\times\hat{J}$ acts as a combination of ladder operator for $\hat{J}$.
  In fact, this doublet does not have direct transitions between $J_{111}=3/2$
  and $J_{111}=-3/2$ states. The Rashba splitting appears only in the third
  order of $\Tilde{\mathcal{F}}$.}
\bibitem{notaShava}
{ Notice that these corrections do not coincide with the ones presented in
  Ref.~\cite{shanavas2014theoretical} because of the different orientation of
  the electric field.}
\bibitem{notaMirror}
{ For a mirror symmetry about the plane $M_{n_\perp}$ we have
  $\Omega(k_{//},k_{\perp})=-\Omega(k_{//},-k_{\perp})$. Therefore, the
  integral over all the BZ of this quantity is zero.}
\end{thebibliography}

\end{document}